\documentclass[10pt,journal,letterpaper]{IEEEtran}
\IEEEoverridecommandlockouts
\usepackage[T1]{fontenc}
\usepackage{aecompl}
\usepackage[numbers,sort&compress]{natbib}
\usepackage{amsmath,amssymb,amsfonts}
\usepackage{amsthm}
\usepackage{float}
\usepackage{booktabs}
\makeatletter
\renewcommand{\maketag@@@}[1]{\hbox{\m@th\normalsize\normalfont#1}}
\makeatother
\usepackage{graphicx}
\usepackage{verbatim}
\usepackage{subcaption}
\usepackage{textcomp}
\usepackage{xcolor}
\usepackage{stfloats}
\usepackage{tabularx}
\usepackage{array}
\usepackage{makecell}
\usepackage{url}
\makeatother
\usepackage{textcomp}
\usepackage[marginal]{footmisc}
\usepackage{comment}
\usepackage{geometry}
\usepackage{setspace}
\allowdisplaybreaks[4]
\usepackage{algpseudocode}
 \usepackage[linesnumbered,ruled]{algorithm2e}

\def\BibTeX{{\rm B\kern-.05em{\sc i\kern-.025em b}\kern-.08em
    T\kern-.1667em\lower.7ex\hbox{E}\kern-.125emX}}

\AddToHook{env/thebibliography/begin}{\interlinepenalty=10000\relax}
\begin{document}
\bstctlcite{IEEEBSTControl}
\title{MATE: Multi-Agent Traffic Engineering for Topology-Aware Distributed SRv6 Scheduling under Bursty Traffic}
\author{Yuang Chen, \IEEEmembership{Member, IEEE}, Shuyi Liu, Chengdi Lu, Fengqian Guo, \\ Hancheng Lu, \IEEEmembership{Senior Member, IEEE}, and Chang Wen Chen, \IEEEmembership{Life Fellow, IEEE}
\thanks{This work was supported in part by the National Natural Science Foundation of China under Grant U21A20452, in part by Hong Kong Research Grants Council (GRF-15213322, GRF-15229423). Yuang Chen was a doctoral student at the Future Network Laboratory, University of Science and Technology of China (USTC), Hefei, P. R. China (e-mail: yuangchen21@mail.ustc.edu.cn), and he is currently a Digital Futures Postdoctoral Fellow at KTH Royal Institute of Technology, Stockholm, Sweden (e-mail: yuangc@kth.se). Shuyi Liu, Chengdi Lu, Fengqian Guo, and Hancheng Lu are with the Laboratory of Future Networks, University of Science and Technology of China (USTC), Hefei, P. R. China (e-mail: \{lsy55, lcd1999\}@mail.ustc.edu.cn; fqguo@ustc.edu.cn; hclu@ustc.edu.cn). Hancheng Lu is also with Deep Space Exploration Laboratory, Hefei 230088, China. Chang Wen Chen is with the Department of Computing, The Hong Kong Polytechnic University, Hong Kong, China (e-mail: changwen.chen@polyu.edu.hk).}}

\maketitle

\begin{abstract}
As network traffic becomes increasingly dynamic and bursty, conventional centralized traffic engineering (TE) suffers from long control-loop delays caused by network-state aggregation, centralized computation, and policy distribution.
In this paper, we propose MATE, a multi-agent TE method for proactive distributed scheduling in hybrid networks with partial deployment of Segment Routing over IPv6 (SRv6).
MATE uses historical traffic stability to characterize the burst risk of each traffic demand and path sensitivity to quantify the ability of candidate paths to accommodate additional traffic, thereby enabling differentiated proactive scheduling tailored to heterogeneous traffic characteristics.
MATE adopts an offline-centralized training and online-distributed execution architecture. 
During training, a shared global critic evaluates joint actions under a reward combining maximum link utilization (MLU) with variability-weighted path sensitivity, while counterfactual difference rewards provide agent-specific credit for actor learning.
During online execution, edge agents generate traffic-splitting policies from local observations and apply them through SRv6, avoiding centralized optimization in each scheduling interval.
Experiments on the China Environment for Network Innovations (CENI) testbed and simulations on the Sanet topology show that MATE outperforms a multi-agent reinforcement learning (MARL) ablation baseline without agent-specific credit assignment, COPE, and TEAL in most evaluated scenarios and approaches centralized FIGRET on measured traffic traces.
On CENI-Abilene, MATE achieves MLU within 10\% of the linear-programming optimum in up to 90.1\% of test instances. MATE also achieves lower median normalized MLU than FIGRET in most tested failure scenarios, with mean neural-inference times of 2.05--3.14~ms.
\end{abstract}


\section{Introduction}

\begin{figure}[t]
\centering
\includegraphics[scale=0.47]{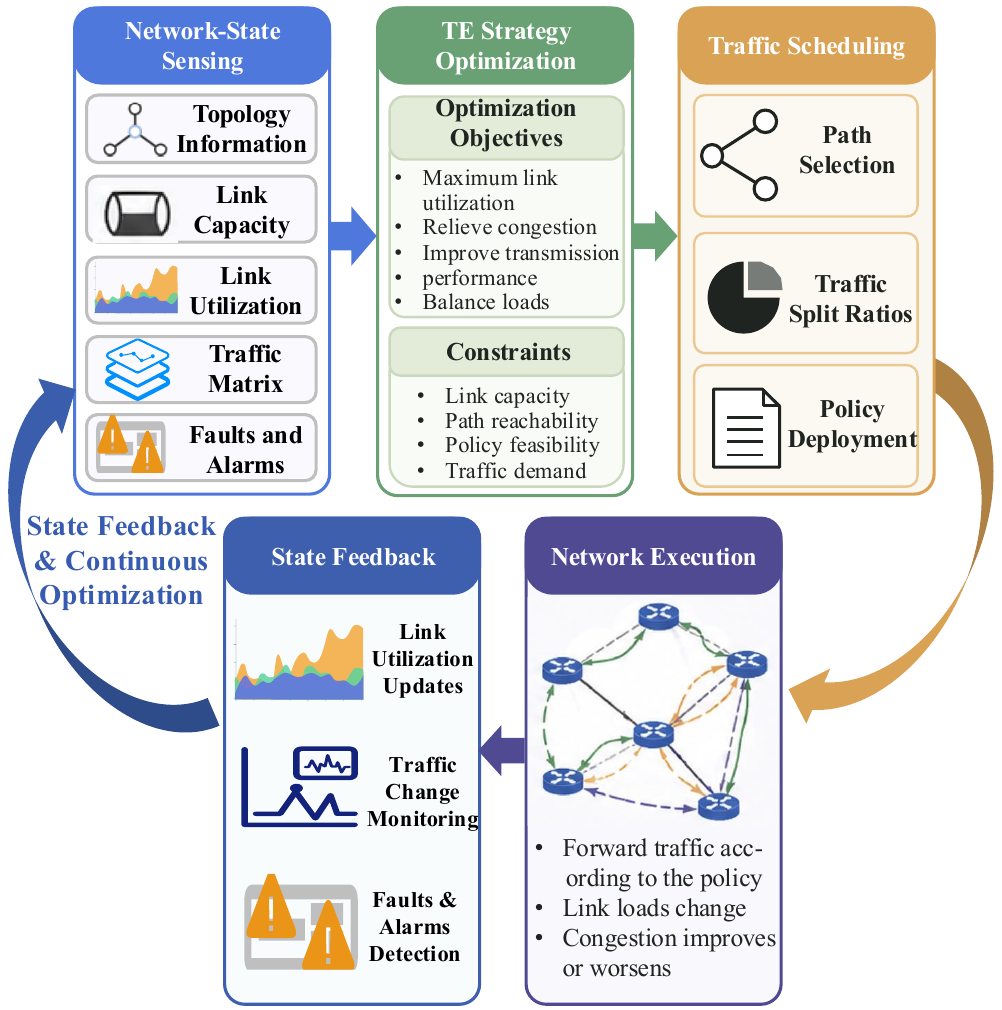}
\caption{\small The Key Procedure of Traffic Engineering.}
\vspace{-1.5em}
\label{fig1}
\end{figure}

\IEEEPARstart{D}{ynamic} and bursty traffic makes timely traffic engineering (TE) increasingly challenging~\cite{figret,teal}.
TE optimizes routing-path selection and traffic-splitting ratios according to network topology, link capacities, traffic demands, and failure states~\cite{teal}.
As illustrated in Fig. ~\ref{fig1}, TE continuously updates these policies using feedback on link utilization and traffic variations, forming a closed-loop process of network-state sensing, strategy optimization, network execution, and state feedback~\cite{rfc9522}.
Demand-aware TE typically optimizes routing policies for measured or predicted demands~\cite{teal,dote}. 
However, sudden demand changes can be difficult to predict accurately, creating discrepancies between the traffic conditions used for optimization and those encountered during policy execution~\cite{dote,figret}.
Traditional centralized TE further amplifies this mismatch due to the control-loop delay~\cite{figret}.
In particular, network-state aggregation, centralized computation, and policy distribution collectively introduce a delay between observing the network and applying the resulting policy~\cite{redte}.
When traffic demands evolve on a timescale comparable to or shorter than the control-loop delay, even rapidly computed policies may be outdated upon deployment~\cite{redte,teal}.
Therefore, effective TE requires both timely online execution and tolerance to demand uncertainty~\cite{redte,figret}.

\begin{figure}[h]
\centering
\includegraphics[scale=0.28]{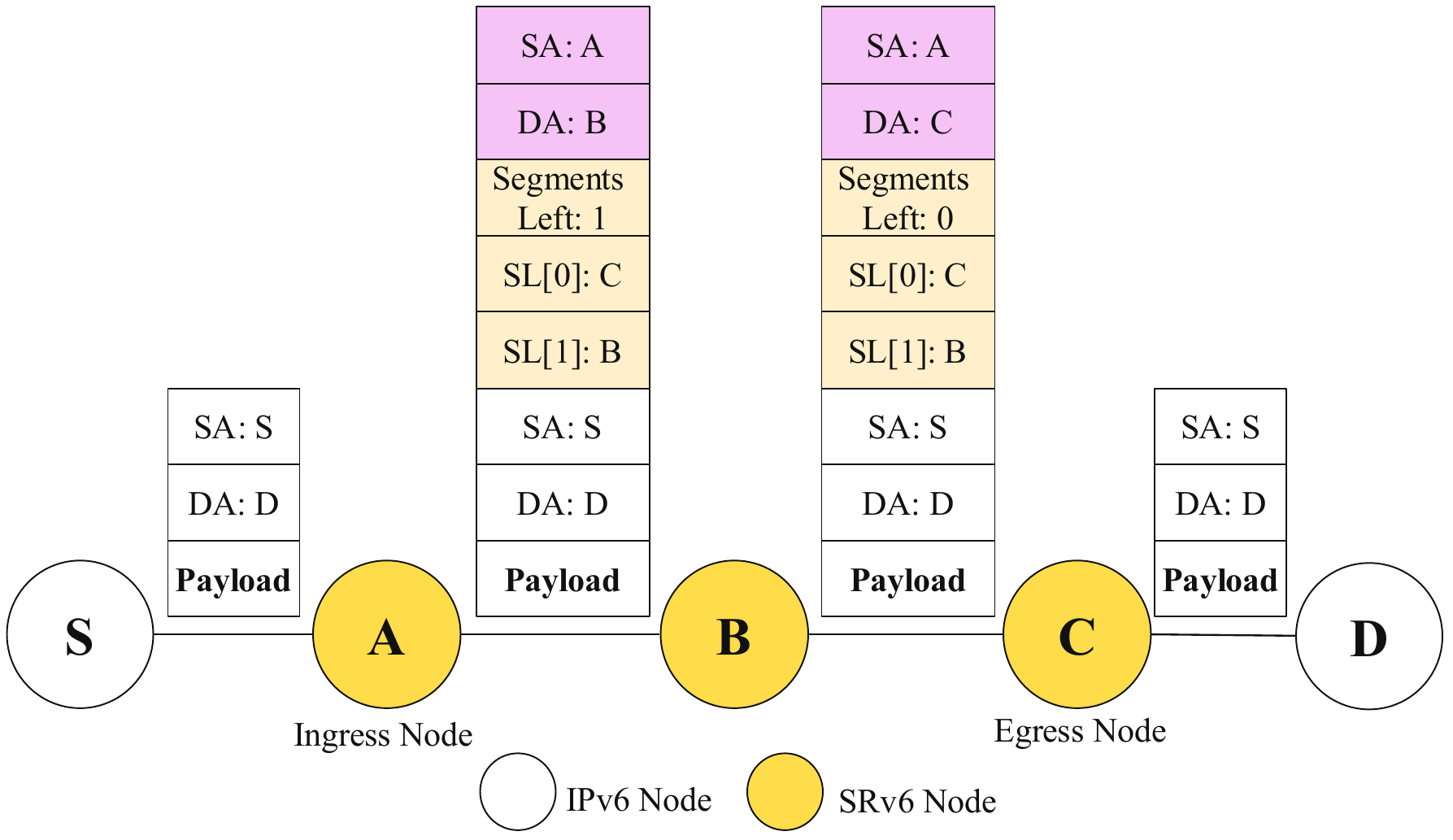}
\caption{\small The transmission of packets over IP/SRv6 networks.}
\vspace{-1.5em}
\label{fig2}
\end{figure}

\par In dynamic networks, segment routing over IPv6 (SRv6) provides a data-plane foundation for flexible traffic engineering by allowing ingress nodes to encode explicit forwarding instructions in packet headers \cite{SRv6,srh}.
As illustrated in Fig. \ref{fig2}, ingress node $\mathbf{A}$ encapsulates a packet with an outer IPv6 header and a segment routing header (SRH), steering it through $\mathbf{B}$ and then $\mathbf{C}$.
Node $\mathbf{B}$ advances the active segment, while egress node $\mathbf{C}$ decapsulates the packet and forwards the original packet to $\mathbf{D}$~\cite{SRv6}.
Ordinary IPv6 transit nodes forward packets toward the active segment identifier (SID) using the outer IPv6 destination address, without processing the SRH~\cite{srh,rfc8402}.
Once candidate paths are provisioned, ingress nodes can directly adjust traffic-splitting policies without reconfiguring every transit device in each scheduling interval~\cite{rfc8402,rfc9256}.
Meanwhile, device heterogeneity, upgrade costs, and operational risks generally necessitate incremental SRv6 deployment, resulting in hybrid networks where conventional IP forwarding and SRv6 coexist~\cite{optimaldeploy,10436774}.
In hybrid IP/SRv6 networks, the feasible paths depend on both SRv6 deployment and the underlying IP routing~\cite{optimaldeploy,rfc8402}.

\par
\begin{figure}[h]
\centering
\includegraphics[scale=0.28]{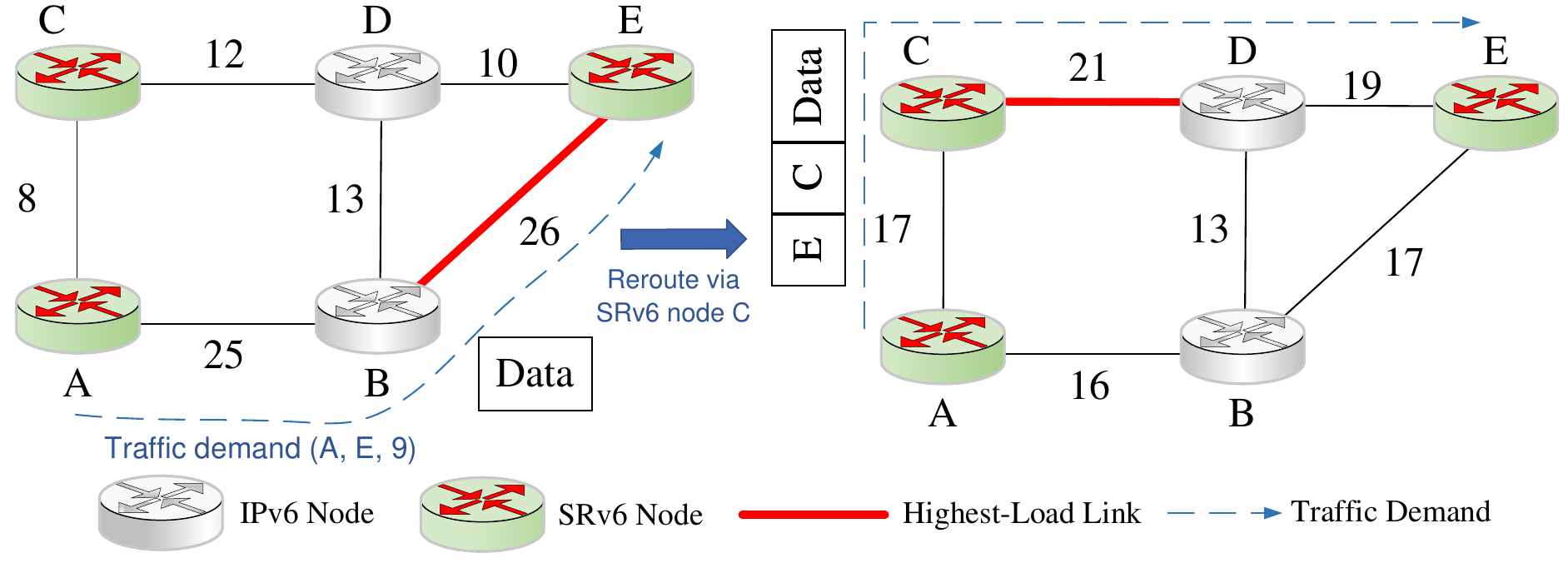}
\caption{\small Traffic rerouting and load balancing in hybrid IP/SRv6 networks.}
\vspace{-0.5em}
\label{fig3}
\end{figure}

 Building on the path-control mechanism described above, SRv6-TE optimizes candidate-path selection and traffic-splitting ratios, with ingress nodes enforcing the resulting policies through segment-list encapsulation~\cite{rfc9256,SRv6}. 
A common optimization objective in TE is to minimize the maximum link utilization (MLU)~\cite{figret}, thereby alleviating load concentration and increasing headroom for demand growth~\cite{klp,dote}.
In Fig.~\ref{fig3}, all links have capacity set to 100 bandwidth units.
Under default IP routing, the red link carries the highest load, resulting in an MLU of 26\%. 
When a TE policy reroutes the demand of 9 bandwidth units from $\mathbf{A}$ to $\mathbf{E}$ over the feasible path $\mathbf{A} \rightarrow \mathbf{C} \rightarrow \mathbf{D} \rightarrow \mathbf{E}$, the traffic bypasses the previously most heavily loaded link, reducing the network-wide MLU to 21\%.
This example illustrates how adjusting traffic allocation through SRv6 path control can improve network load distribution. 
The routing policy can be computed centrally or generated locally, with ingress SRv6 nodes enforcing the resulting traffic allocation~\cite{rfc8402}.

\par Distributed online decision-making can shorten TE control-loop delays by allowing routers to generate decisions locally~\cite{redte}.
Motivated by this principle, this work develops an architecture in which a central controller operates in conjunction with edge SRv6 nodes.
We separate centralized path provisioning and offline learning from distributed online traffic splitting.
The controller computes and updates feasible candidate paths, establishes SRv6 tunnels, maintains network information, handles failures, and trains agents.
During routing operation, each ingress agent maps local traffic history, link utilization, and capacities to traffic-splitting decisions, enforced through SRv6 encapsulation~\cite{SRv6,rfc9256}.
Therefore, routing decisions avoid network-wide state aggregation and centralized optimization, while provisioning, maintenance, and model updates retain controller support.
Nevertheless, decisions at different ingress nodes remain coupled because their traffic may share physical links, making coordination necessary for network-wide
optimization~\cite{teal,guodistri}.

\par However, SRv6 path programmability and ingress-based policy execution alone cannot guarantee TE robustness under bursty traffic.
When future traffic demands are difficult to predict accurately, TE policies optimized solely for current or predicted demands may overload links if actual demands deviate substantially from those assumed during optimization~\cite{dote}.
Conversely, uniformly conservative protection against potential bursts may distribute stable traffic across more, and potentially longer, paths, increasing resource consumption under normal conditions~\cite{figret}.
Therefore, balancing network efficiency under normal conditions against robustness to burst-induced congestion remains a central challenge for dynamic TE~\cite{cope}.
In the distributed setting considered in this work, this balance must be achieved through coordinated decisions based on local online observations, while accounting for interactions among demands sharing network links~\cite{teal,guodistri}.

\vspace{-1em}

\subsection{Motivation and Challenges}

\par Existing TE approaches for bursty traffic broadly follow two technical directions. 
The first adopts reactive scheduling, which detects network-state changes after a traffic burst and then recomputes and deploys TE policies.
For example, RedTE~\cite{redte} develops real-time distributed TE for sub-second traffic bursts.
The effectiveness of such approaches hinges on minimizing the control-loop delay associated with state monitoring, policy computation, and configuration deployment.
As network size and traffic demand increase, information collection, centralized computing, and rule deployment can introduce additional latency, making it difficult for TE to consistently keep pace with the evolution of traffic bursts. 
The second adopts robust or proactive scheduling, which explicitly accounts for future demand uncertainty when generating policies.
Demand-oblivious routing, COPE~\cite{cope}, together with path sensitivity-based robust TE fall into this category~\cite{makingintra,cope,semi,jupiter,enabling2}. 
These approaches illustrate different ways of balancing demand awareness and protection against uncertainty.
However, uniform worst-case protection can impose unnecessary path diversity and detours on stable demands~\cite{figret}.
Since historical traffic variability differs across source--destination pairs, proactive protection should be stronger for burst-prone demands while limiting additional overhead for stable traffic~\cite{figret}.

\par FIGRET~\cite{figret} provides demand-specific robustness by tailoring constraints to historical traffic variability.
However, the utilization impact of an unexpected demand increase also depends on path capacities and competing traffic on shared links~\cite{dote,teal}.
Although FIGRET~\cite{figret} incorporates topology and capacity information into its training objective, its neural policy primarily maps historical traffic matrices (TMs) to routing configurations without taking current topology features as inference inputs~\cite{figret}.
This motivates conditioning distributed policies on both traffic history and locally available link-state features.

\par The aforementioned observations motivate a TE  approach that jointly considers traffic burstiness and candidate-path capacity to provide differentiated proactive protection. The challenge is to support such protection under local online observations, leading to the following research questions.

\par \textbf{Challenge 1: How can traffic burst risk and candidate-path capacity be jointly considered for differentiated proactive scheduling under local online observations?}

\par Addressing this challenge requires identifying the historical stability of individual demands and relating their burst risk to the capacity and topological properties of their candidate paths~\cite{figret}.
An effective policy should provide stronger protection for burst-prone demands while avoiding unnecessarily conservative restrictions on stable demands.
Beyond accounting for burst risk, a more fundamental question concerns where these TE decisions should be generated. 

\par Centralized TE jointly optimizes traffic allocation using network-wide information ~\cite{dote,teal}, but its state collection, centralized computation, and policy deployment can delay policy application~\cite{redte}.
Distributed TE instead decomposes network-wide optimization into local decision problems at network nodes~\cite{guodistri,redte}.
SRv6 ingress nodes enforce them over preprovisioned paths through SR policy steering and source encapsulation~\cite{rfc9256,SRv6}.
This design removes the need for network-wide state
aggregation and centralized policy generation in each
scheduling interval, providing a basis for a shorter
online control loop.
However, each ingress node has access only to local observations, while traffic-splitting decisions at different nodes are coupled through links shared by their candidate paths.
A locally beneficial traffic shift may increase loads on shared links elsewhere in the network~\cite{teal}.
Consequently, independently generated local decisions do not necessarily minimize the network-wide MLU~\cite{guodistri}.
This leads to the second challenge.

\par \textbf{Challenge 2: How can fast distributed execution be achieved using only local online observations at edge nodes, while maintaining performance close to that of centralized global TE optimization?}

\par Learning coordinated policies for distributed execution also introduces a multi-agent credit assignment challenge.
Network-wide MLU depends on the joint actions of all edge agents, whereas a shared global reward does not directly reveal each agent's contribution to a performance change~\cite{coma}.
This ambiguity can hinder coordinated learning by making the effects of individual scheduling decisions difficult to distinguish.
Therefore, the training mechanism must connect local policy updates to their effects on the global TE objective.
This motivates the third challenge.

\par \textbf{Challenge 3: How can coordination among multiple edge agents be strengthened under local observations and distributed actions, so that their joint behavior aligns with global TE optimization
objectives?}



\par Addressing this challenge requires agent-specific learning signals that assess each scheduling decision in the context of the other agents' actions.
Counterfactual comparisons can provide such credit by varying one agent's action while holding the others fixed, helping relate individual decisions to the shared TE objective~\cite{coma,guodistri}. Coordination should be established during centralized training and retained in the learned policies, enabling distributed online decisions based solely on local observations without inter-agent exchange of observations or actions~\cite{maddpg}.

\vspace{-0.8em}

\subsection{Main Contributions}

\par To address these challenges, we propose MATE, a multi-agent TE method for distributed SRv6 scheduling.
MATE jointly considers traffic variability and candidate path topology, using traffic stability and path sensitivity to characterize burst risk and paths' ability to accommodate additional traffic, respectively. 
We formulate scheduling as network-wide MLU minimization with stability-dependent path sensitivity constraints, enabling differentiated proactive protection and steering burst-prone demands toward higher-capacity paths.
MATE adopts offline centralized training and online distributed execution.
By avoiding repeated network-wide state aggregation and policy distribution, MATE aims to combine fast distributed response with congestion-control performance approaching that of state-of-the-art centralized TE. 
The main contributions are summarized as follows:

\begin{itemize}
    \item We formulate differentiated proactive SRv6-TE using historical traffic stability and capacity-aware path sensitivity.
    The formulation strengthens protection for burst-prone demands while limiting unnecessary conservatism for stable traffic.
    
    \item We propose MATE with offline centralized training and online distributed execution, decomposing network-wide TE into local decisions at edge SRv6 nodes.
    Each agent generates splitting decisions from local observations and applies them through ingress encapsulation, reducing reliance on centralized online optimization.
    
    \item We incorporate agent-specific counterfactual credit assignment into centralized training for distributed SRv6-TE. 
    A shared global critic compares joint actions with alternatives that vary one agent's decision, producing difference rewards to guide coordinated policy learning under the congestion-and-sensitivity objective.

    \item We systematically evaluate MATE on a WAN testbed built on the China Environment for Network Innovations (CENI)~\cite{ceni} and through simulations on the Sanet topology from TopologyZoo~\cite{topozoo}, modeled with 43 nodes and 90 links.
    MATE outperforms the MARL ablation baseline, COPE~\cite{cope}, and TEAL~\cite{teal} in most tested scenarios and approaches centralized FIGRET~\cite{figret} on real-world traffic traces while using distributed execution. On synthetic traces, MATE remains competitive with DOTE~\cite{dote}.
    On CENI-Abilene, MATE keeps normalized MLU at or below $1.1$ in up to $90.1\%$ of test instances.
    It also achieves lower median normalized MLU than FIGRET in most evaluated link-failure scenarios, with mean online inference times of $2.05$--$3.14$~ms.
\end{itemize}

\par The remainder of this paper is organized as follows.
Sec.~II reviews related work.
Sec.~III presents the hybrid IP/SRv6 system model and the burst-aware distributed TE.
Sec.~IV describes MATE's design, centralized training, and distributed execution.
Sec.~V evaluates its performance, robustness, and online inference time through CENI testbed experiments and simulations.
Sec.~VI concludes the paper and outlines future directions.

\vspace{-1em}

\section{Related Work}

\subsection{SRv6 and Learning-Based Traffic Engineering}

\par TE optimizes routing paths and traffic-splitting ratios according to network topology, link capacities, and traffic demands~\cite{teal}.
Traditional approaches usually formulate TE as linear programming (LP), integer LP (ILP), or mixed-ILP (MILP) problems and solve them using general-purpose optimizers~\cite{cope,moreno2017srte,tm,klp}.
Li and Yeung~\cite{klp}, for example, formulate segment-routing path selection as an MILP to minimize MLU. 
These optimization-based methods can provide high-quality or optimal solutions for small networks or slowly changing conditions~\cite{moreno2017srte,tm,klp}.
However, their computational burden can become substantial as the network, demand set, and candidate path set grow, particularly when integer routing decisions are involved, limiting their suitability for time-sensitive online operation~\cite{teal,klp}.

\par To reduce online optimization overhead, approximation, heuristic, and learning-based TE methods have been developed~\cite{q-sr,midpoint25,testt,enero,teal}.
Q-SR~\cite{q-sr} provides an extensible optimization
framework with offline approximation and online
primal--dual algorithms for segment routing.
Heuristic methods rely on hand-designed search or
selection rules, and their performance can depend on
parameter choices, search budgets, and traffic
characteristics~\cite{midpoint25,testt,enero,teal}.
By comparison, deep reinforcement learning (DRL) learns scheduling policies through interaction and generates online decisions through neural inference ~\cite{testt,enero,teal,redte}.
Nevertheless, their deployment performance depends on how well the learned policies generalize to the traffic and network conditions encountered during
execution~\cite{teal,telgen}.

\par Several studies demonstrate the effectiveness of learning for TE. 
ENERO~\cite{enero} combines DRL with local search for real-time WAN routing optimization.
Graph neural networks (GNNs) have also been used to capture dependencies among topology, routing paths,
and traffic demands~\cite{routenet,teal}.
TEAL~\cite{teal} encodes connectivity, link capacities, and traffic demands using a flow-centric GNN, generates allocations through multi-agent DRL, and refines them with a few iterations of the alternating direction method of multipliers (ADMM).
DOTE~\cite{dote} directly maps historical TMs to routing configurations for the next interval, replacing online optimization with model inference without explicitly predicting future demands.

\par These works show that learning can reduce online optimization costs and generate high-quality routing decisions. 
Nevertheless, faster decisions alone do not ensure robustness to unpredictable traffic bursts~\cite{figret}.
This motivates proactive scheduling that tolerates future traffic uncertainty, beyond accelerating route computation.

\vspace{-0.8em}

\subsection{Robust and Proactive Traffic Engineering}

\par Robust TE seeks to maintain network performance under uncertain traffic demands~\cite{makingintra,cope,semi,figret}.
Early demand-oblivious routing computes configurations independently of the current TM while providing congestion guarantees~\cite{azar2003oblivious,makingintra}.
COPE~\cite{cope} optimizes performance for expected demands while retaining a worst-case performance guarantee. 
Semi-Oblivious TE~\cite{semi} combines demand-independent path selection with demand-aware traffic allocation, balancing robustness and routing efficiency.
However, stronger worst-case protection can compromise efficiency under normal traffic conditions~\cite{cope,semi,jupiter,figret}.
In particular, uniformly conservative constraints may unnecessarily spread stable demands across additional or longer paths, increasing resource consumption even when only a subset of demands is highly bursty~\cite{jupiter,figret}.

\par To reduce this conservatism, subsequent studies incorporate traffic characteristics into path or topology optimization to improve robustness~\cite{jupiter,enabling2,figret}.
In reconfigurable data centers, Jupiter combines traffic and topology engineering with path diversity and variable hedging to tolerate prediction errors~\cite{jupiter}.
COUDER jointly optimizes topology and routing over historical traffic scenarios and reduces routing sensitivity to outlier demands, limiting the need for frequent topology reconfiguration~\cite{enabling2}.
LARRI~\cite{larri} combines supervised learning and graph neural networks to learn adaptive range routing from historical demands, addressing the trade-off between routing efficiency and robustness to future traffic variations.
More recent work differentiates protection according to the variability of individual demands. For example, FIGRET~\cite{figret} provides demand-specific robustness by tailoring path-sensitivity constraints to historical traffic variability.
Its formulation accounts for path capacity, while its neural policy uses historical TMs without explicit current-topology inputs.
This work focuses on combining differentiated proactive protection with distributed SRv6 execution based on local traffic histories and link-state observations.

\vspace{-1em}

\subsection{Distributed and Multi-Agent Traffic Engineering}

\par Distributed TE moves routing decisions to network nodes to reduce dependence on centralized online control~\cite{guodistri,redte}.
HALO~\cite{halo} iteratively adjusts hop-by-hop traffic splitting using disseminated link-state information.
Learning-based approaches further enable cooperative
routing policies.
Guo et al. in~\cite{guodistri} develop a multi-agent TE framework for hybrid software-defined networks (SDN), using difference rewards to coordinate agents with partial observations.
RedTE~\cite{redte} combines local online inference with centralized cooperative training to mitigate subsecond traffic bursts, emphasizing the importance of reducing the complete control-loop delay.

\par Recent studies explore richer network representations and additional information for routing decisions.
Aether~\cite{11044607} employs elastic multi-agent graph transformers to improve generalization across network environments.
DITE~\cite{11023120} supplements local observations with sparse network-wide guidance and combines imitation learning with reinforcement learning.
PathGNN~\cite{pathgnn} learns from optimal routing
solutions through supervised training and uses
path--link graph modeling with inter-router message
exchanges to adapt to traffic fluctuations and link
failures.
These approaches illustrate different trade-offs between routing performance and online information requirements.

\par Coordination remains essential because traffic-splitting decisions at different nodes interact through shared links~\cite{guodistri,redte}.
In decentralized WAN control, Symphony~\cite{devraj2025stable} regularizes TE objectives to stabilize allocations under inconsistent demand estimates across autonomous controllers.
In multi-agent learning, a shared global reward can obscure the contributions of individual actions. COMA~\cite{coma} addresses this credit assignment problem through a centralized critic and a counterfactual baseline that varies one agent's action while holding the others fixed, providing a foundation for cooperative policy learning.

\par MATE focuses on variability-aware traffic splitting over paths realizable under partial SRv6 deployment. 
It combines local traffic histories and link-state observations with a variability-weighted path-capacity objective. 
This differs in design emphasis from link-level splitting in hybrid SDN~\cite{guodistri} and adaptive range routing~\cite{larri}.
MATE combines centralized actor--critic training~\cite{maddpg} with critic-based counterfactual credit assignment~\cite{coma,guodistri} to coordinate distributed ingress policies under a shared congestion-and-sensitivity
objective.


\section{System Model}

\subsection{Scenario Description}

\par Table \ref{tab1} summarizes the main notation. 
We model the hybrid IP/SRv6 network as a graph $G = (\mathcal{N}, \mathcal{L})$, where $\mathcal{N}$ and $\mathcal{L}$ denote the node and link sets, respectively. 
Link $l$ has capacity $C(l)>0$. The nominal bottleneck capacity of path $p$ can be defined as

\begin{equation}\label{eq1}
    C_p = \min_{l \in p} C(l).
\end{equation}

\par Let $D_t$ denote the traffic matrix (TM) at time $t$.
For each source–destination (SD) pair $(s,d)$, traffic demand $f_{sd}$ has bandwidth requirement $d_{sd}$ and is split over a predefined candidate path set $P_{sd}$.
To limit the routing decision space and the amount of routing state, $P_{sd}$ contains only $K$ candidate paths selected before traffic transmission, rather than all feasible paths between $s$ and $d$. 
These candidate paths are computed using a $K$-shortest-path algorithm, with preference given to paths with minimal link overlap to increase path diversity and improve resilience to link failures. Each path admitted to $P_{sd}$ must have an SRv6 segment-list representation that realizes the path under the given deployment and underlying IP routing~\cite{gofor}.

\begin{table}[h]
    \centering
    \caption{Summary of Main Notation}
    \label{tab1}
    \footnotesize
    \renewcommand{\arraystretch}{1.15}
    \setlength{\tabcolsep}{4pt}
    \begin{tabularx}{\columnwidth}{
        @{}
        >{\centering\arraybackslash}p{0.27\columnwidth}
        >{\raggedright\arraybackslash}X
        @{}
    }
        \toprule
        \textbf{Symbol} & \textbf{Description} \\
        \midrule
        $G=(\mathcal{N},\mathcal{L})$
        & Network topology graph, with node set $\mathcal{N}$
          and link set $\mathcal{L}$. \\

        $\mathcal{N}^{1},\mathcal{N}^{2},N,M$
        & Sets of edge and intermediate nodes, with cardinalities
          $N$ and $M$, respectively. \\

        $P_{sd}$
        & Set of candidate paths from source $s$ to destination $d$. \\

        $C(l),C_p$
        & Bandwidth capacities of link $l$ and path $p$, respectively. \\

        $D_t$
        & Traffic matrix (TM) at time $t$. \\

        $f_{sd},d_{sd}$
        & Traffic demand from $s$ to $d$ and its magnitude, respectively. \\

        $g_p$
        & Fraction of a demand assigned to path $p$. \\

        $z_{l,p}$
        & Binary indicator equal to 1 if link $l$ belongs to path $p$,
          and 0 otherwise. \\

        $U_{\max}$
        & Maximum link utilization (MLU) in the network. \\

        $d'_{sd},\delta_{sd}$
        & Predicted demand and prediction error, with
          $\delta_{sd}=d_{sd}-d'_{sd}$. \\

        $S_p^{sd}$
        & Path sensitivity, defined as $S_p^{sd}=g_p/C_p$. \\

        $\Theta(s,d)$
        & Strictly positive path-sensitivity limit for $f_{sd}$; smaller values impose stronger protection for less stable demands. \\

        $\alpha,K$
        & SRv6 deployment ratio among intermediate nodes and number of candidate paths per demand. \\

        $w,\tau$
        & Traffic-history window and target-network soft-update coefficient. \\

        $H$
        & Number of complete counterfactual actions sampled per agent and evaluated state. \\

        $\mathrm{CV}_{id},A_t$
        & Historical coefficient of variation of demand $(i,d)$ and unnormalized sensitivity penalty at time $t$. \\

        $r_t$ & Shared global reward combining MLU and variability-weighted path sensitivity.\\

        $R_t^i$ & Counterfactual difference reward providing agent-specific credit for actor training. \\

        $\operatorname{sg}(\cdot)$ & Stop-gradient operation, treating its argument
as constant during differentiation.\\
        \bottomrule
    \end{tabularx}
\end{table}

\par We consider a hybrid IP/SRv6 network in which the node set $\mathcal{N}$ is partitioned into the edge node set $\mathcal{N}^{1}$ and the intermediate node set $\mathcal{N}^{2}$, with cardinalities $N$ and $M$, respectively.
All edge nodes are SRv6-enabled, and the deployment ratio $\alpha$ denotes the fraction of intermediate nodes that are SRv6-enabled.
A centralized controller supports candidate path computation and updates, SRv6 tunnel establishment, network failure detection, and maintenance of the network topology and resource information.
Edge nodes connect the network to external networks
and handle incoming and outgoing traffic.
Their primary responsibilities include traffic classification, policy binding, traffic splitting, and SRv6 encapsulation and decapsulation at ingress and egress, respectively.
Intermediate nodes provide transit forwarding: SRv6-enabled nodes execute the corresponding segment behaviors when acting as segment endpoints, while conventional IPv6 nodes forward packets according to the outer IPv6 destination address.

\par
\begin{figure}[h]
\centering
\includegraphics[scale=0.28]{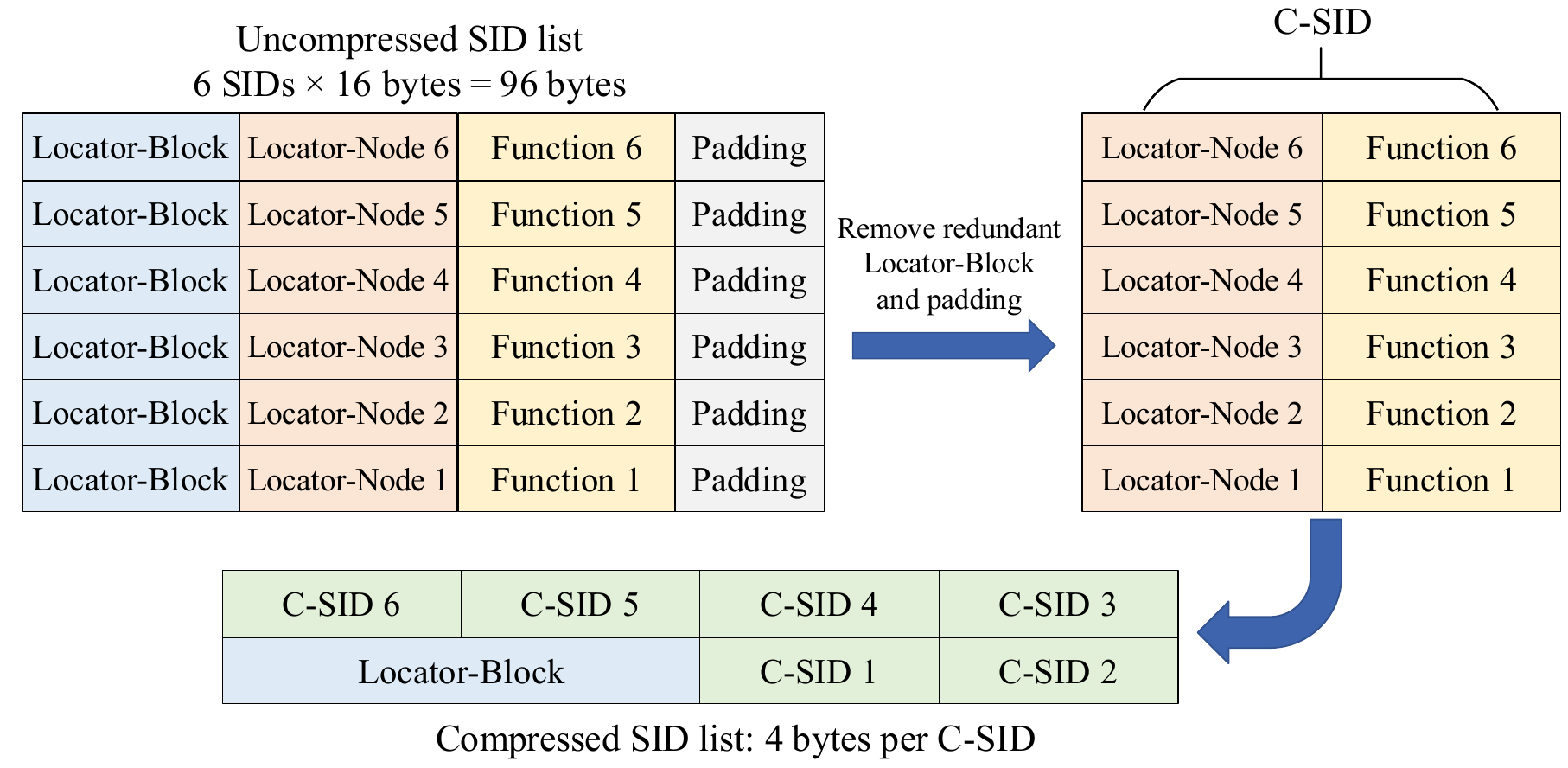}
\caption{\small The SRH compression mechanism.}
\vspace{-2em}
\label{fig4}
\end{figure}

 SRv6 encapsulation adds header overhead~\cite{11303321}. 
RFC 9800~\cite{rfc9800} provides compressed segment-list encoding as an option for reducing it.
Fig. \ref{fig4} illustrates a 32-bit (i.e., 4-byte) compressed segment identifier (C-SID) format with a common Locator-Block and per-segment Locator-Node and Function fields. 
Removing repeated blocks and redundant padding reduces each segment’s representation from 16 to 4 bytes, in addition to the shared block and container overhead.
During forwarding, SRv6 segment endpoints combine the common block with the next C-SID to form the active IPv6 destination address \footnote{This encoding option is deployment background; the reported evaluation does not separately quantify compression gains.}.


\vspace{-1em}

\subsection{Problem Formulation}\label{sec:mate-problem-formulation}

\par Let $\mathcal{Q} = \{(s,d)\in\mathcal{N}^{1} \times \mathcal{N}^{1}: s\neq d\}$ denote the set of source–destination pairs. 
For each demand $f_{sd}$ with bandwidth requirement $d_{sd} \geq 0$, the ingress node $s$ distributes its traffic over the candidate paths in $P_{sd}$. 
Let $g_p \in \left[0,1\right]$ denote the fraction assigned to path $p \in  P_{sd}$, with its dependence on $(s,d)$ omitted for brevity. 
The splitting ratios must satisfy

\begin{equation}\label{eq2}
    \sum_{p\in P_{sd}} g_p = 1,
    \qquad \forall (s,d) \in \mathcal{Q}.
\end{equation}

\par This constraint ensures that each demand is completely allocated across its candidate paths.

\par For each link $l \in \mathcal{L}$, let $z_{l,p}\in\{0,1\}$ be a fixed indicator equal to one if path $p$ traverses link $l$ and zero otherwise.
Given the link capacity $C(l)>0$, the offered load on link $l$ is normalized by its capacity.
We introduce $U_{\max}\geq 0$ as a common upper bound on these normalized link loads, as follows:

\begin{equation}\label{eq3}
    \frac{1}{C(l)}
    \sum_{(s,d)\in\mathcal{Q}} \ \sum_{p\in P_{sd}}
    z_{l,p} \cdot d_{sd} \cdot g_p
    \leq U_{\max},
    \qquad \forall l\in\mathcal{L}.
\end{equation}

\par Minimizing this common upper bound minimizes network-wide MLU.
The resulting optimization problem can be formulated as follows:

\begin{subequations}
\label{eq4}
\begin{align}
    \mathcal{P}1: \quad \underset{\mathbf{g},U_{\max}}{\operatorname{min}}
    \quad & U_{\max},
    \label{eq4a}\\
    \operatorname{subject\ to:}\quad
    & \eqref{eq2},\ \eqref{eq3},
    \label{eq4b}\\
    & 0\leq g_p\leq 1,\quad U_{\max}\geq 0.
    \label{eq4c}
\end{align}
\end{subequations}
where $\mathbf{g} =
\{g_p:(s,d)\in\mathcal{Q},\,p\in P_{sd}\}$ collects all traffic-splitting variables,
and the bounds on $g_p$ in~\eqref{eq4c} apply to every candidate path of each SD pair.
The variable $U_{\max}$ is optimized jointly with $\mathbf{g}$.
At optimality, it equals the maximum offered-load-to-capacity ratio across all links. 
A value above one indicates that the offered load exceeds capacity on at least one link.

\par In practice, traffic measurement, decision computation, and policy installation incur control-loop delay. Consequently, a scheduling decision must be generated before the demand for the next interval is fully observed. Let $d^{\prime}_{sd}$ denote the predicted demand and $\delta_{sd}$ the prediction error, such that 

\begin{equation}\label{eq5}
    d_{sd} = d^{\prime}_{sd} + \delta_{sd}.
\end{equation}

\par Substituting \eqref{eq5} into \eqref{eq3} expresses the link-load constraints in terms of predicted demands and prediction errors, yielding:

\begin{equation}\label{eq6}
\begin{aligned}
    &\frac{1}{C(l)}
    \sum_{(s,d)\in\mathcal{Q}} \ \sum_{p\in P_{sd}} z_{l,p} \cdot g_p \cdot  \bigl(d^{\prime}_{sd} + \delta_{sd} \bigr) \leq U_{\max}, \forall l\in\mathcal{L},
\end{aligned}
\end{equation}
where $d^{\prime}_{sd}$ denotes the predicted demand, and $\delta_{sd}=d_{sd} - d^{\prime}_{sd}$ is the signed prediction error.
The splitting ratio $g_p$ determines the fraction of demand $f_{sd}$ assigned to path $p$ and therefore scales both the predicted load and the additional load associated with the prediction error.


\par Prediction errors are generally smaller for stable demands than for highly variable demands, whose bursts can produce substantial positive deviations. Since $\delta_{sd}$ is unknown until demand is observed, scheduling should limit the extra load caused by unexpected demand growth. Both load components in \eqref{eq6} are scaled by $g_p/C(l)$. Thus, the congestion impact of a prediction error depends on the routing decision and the capacities of the traversed links. To characterize this dependence at the path level, recall that $C_p = \min_{l \in p}C(l)$ is the bottleneck capacity of path $p$. For an individual demand--path contribution, the bottleneck-normalized load can be decomposed as

\begin{equation}\label{eq7}
    \frac{d_{sd} g_p} {C_p}
    =\frac{d^{\prime}_{sd} g_p}{C_p}
    +\frac{\delta_{sd}g_p}{C_p}.
\end{equation}

\par Equation \eqref{eq7} separates the predicted-load contribution from the error contribution $\delta_{sd}g_p/C_p$.
For the same split and positive error, a lower-capacity path experiences a larger increase in normalized load. MATE uses this relationship to motivate capacity-aware protection, without explicitly predicting next-interval demands. Following prior work~\cite{figret,enabling2}, we define the path sensitivity as


\begin{equation}\label{eq8}
    S_{p}^{sd} = \frac{g_p}{C_p},
\end{equation}
where a larger $C_p$ or smaller $g_p$ reduces sensitivity. The additional normalized load is $\delta_{sd} \cdot S_p^{sd}$, linking demand uncertainty to the routing decision and path capacity.

\par A prescribed per-demand limit controls this contribution:

\begin{equation}\label{eq9}
    S_p^{sd} \leq \Theta(s,d), \forall(s,d)\in\mathcal{Q},\quad p\in P_{sd},
\end{equation}
where $\Theta(s,d)>0$ is a prescribed path-sensitivity limit.
Smaller limits represent stronger protection for demands with greater historical variability, while larger limits allow more allocation flexibility for stable demands.

\par Since \eqref{eq9} is equivalent to $g_p \leq \Theta(s,d) C_p$, a path with a larger bottleneck capacity can carry a larger fraction of a demand under the same sensitivity limit. 
For these per-demand constraints to admit a nonnegative unit-sum split, the threshold must satisfy $\Theta(s,d)\sum_{p\in P_{sd}}C_p\geq1$.
These sensitivity limits complement the aggregate link-load constraints in~\eqref{eq3}, which account for contention among demands sharing physical links.

\par For a given traffic realization and prescribed sensitivity thresholds, the sensitivity-constrained scheduling problem can be denoted as:

\vspace{-1.5em}

\begin{subequations}\label{eq10}
\begin{align}
    \mathcal{P}2:\quad
    \underset{\mathbf{g},U_{\max}}
    {\operatorname{minimize}}
    \quad & U_{\max}
    \label{eq10a}\\
    \operatorname{subject\ to}\quad
    & \eqref{eq2},\ \eqref{eq6},\ \eqref{eq9},
    \label{eq10b}\\
    & 0\leq g_p\leq 1,\quad U_{\max}\geq 0.
    \label{eq10c}
\end{align}
\end{subequations}
where $g_p$ and $U_{\max}$ denote the decision variables.
Given realized demands and fixed candidate paths, capacities, and sensitivity thresholds, $\mathcal{P}2$ is an LP that serves as a
sensitivity-constrained reference model.
This formulation describes the balance between network-wide load balancing and per-path sensitivity control.
However, online decisions must be generated before next-interval demands are fully observed.
Therefore, MATE learns splitting policies offline and uses them to generate decisions from local observations.
The learning objective below combines MLU minimization with a soft regularizer that accounts for demand variability and path capacity.

\section{Algorithm Design of MATE}

\par To reduce the overhead of centralized online optimization, MATE distributes TE decisions among edge SRv6 nodes.
Each agent uses a fully connected actor network to map local traffic histories and link-state features to continuous splitting ratios.
DRL learns these policies through interaction without labeled optimal routing solutions, making it suitable for scheduling under bursty traffic~\cite{redte}.
Burst protection is incorporated through the stability-aware path-sensitivity mechanism.
A shared global reward can obscure individual agents' contributions, complicating coordination toward network-wide MLU minimization~\cite{coma,guodistri}.
Therefore, MATE builds on the multi-agent deep deterministic policy gradient (MADDPG) framework~\cite{maddpg}, which extends the model-free actor--critic algorithm DDPG~\cite{ddpg} to multi-agent continuous control.
A shared global critic evaluates joint actions under the congestion-and-sensitivity reward.
MATE further incorporates counterfactual credit assignment~\cite{coma,guodistri}, deriving agent-specific difference rewards by comparing each agent's action with sampled alternatives while holding the others fixed.
These signals guide actor learning toward the shared network-wide objective.

\begin{figure}[h]
\centering
\includegraphics[scale=0.21]{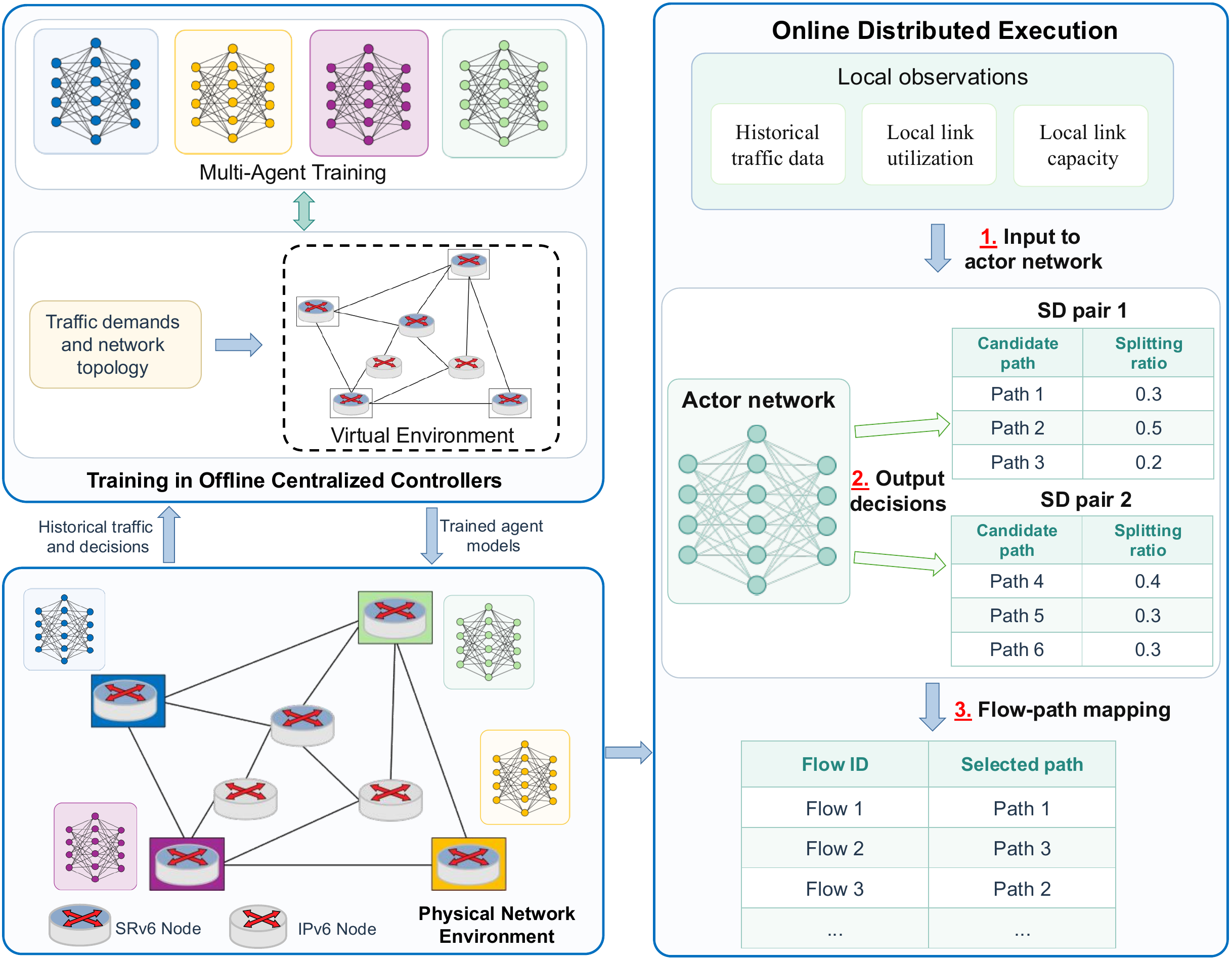}
\caption{\small MATE workflow.}
\vspace{-1em}
\label{fig5}
\end{figure}

\par As illustrated in Fig. \ref{fig5}, MATE combines offline centralized training with online distributed execution.
The controller collects historical network-wide TMs and scheduling experience to train edge-specific policies in a virtual network environment.
Through repeated interaction, agents learn mappings from local observations to splitting decisions, with each agent maintaining its own policy network.
These networks need not have identical architectures.
The trained models are deployed at edge nodes for local online inference without collecting global state for each scheduling decision.
Nodes periodically report newly observed traffic demands and receive updated policy models from the controller.

\begin{figure}[h]
\centering
\includegraphics[scale=0.21]{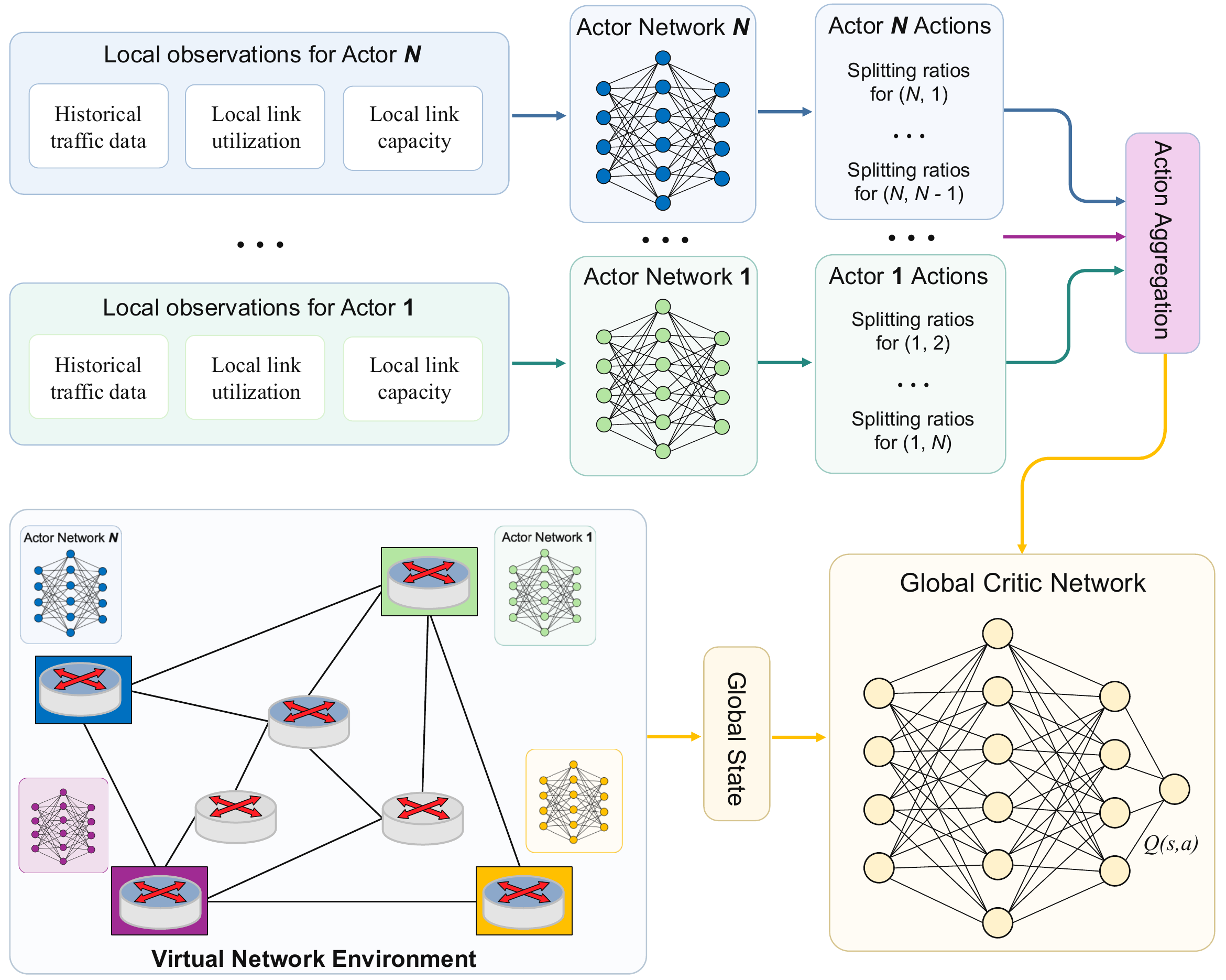}
\caption{\small The offline centralized training of MATE algorithm.}
\vspace{-2em}
\label{fig6}
\end{figure}

\subsection{Offline Training Phase}

\par MATE trains its agents in a virtual network environment using historical traffic data.
This environment supports controlled replay and centralized training with network-wide information.
The training setup is described below.



\subsubsection{\textbf{Local Observation $o_{t}^{i}$}} Each agent $i$ selects its action from its local observation $o_{t}^{i}$.
We use $i$ to denote both an agent and its associated edge node; for this agent, the source index $s$ in Sec.~III corresponds to $i$. The observation contains:

\begin{itemize}
    \item \textbf{Recent traffic history}: To limit input dimensionality, agent $i$ uses demands from its edge node to the other edge nodes over the preceding $w$ TMs:
        \begin{equation}\label{eq11}
            d_t^{i} = \left\{\left(D_{t-w}^{\rm{id}}, \cdots, D_{t-1}^{\rm{id}}\right), d \in \mathcal{N}^{1} \setminus \{i\}\right\},
        \end{equation}
        where $D_\tau^{id}$ is the demand from node $i$ to node $d$ in TM $D_\tau$.

    \item \textbf{Local link utilization:} The utilization of links locally observable at edge node $i$ is $u_t^i = \left\{ u_t^{i,l} \,\middle|\,l \in \mathrm{Adj}_i \right\}$, where $\mathrm{Adj}_i$ is the set of links locally observable at edge node $i$.
    \item \textbf{Local link capacities:} The capacities of these links are represented as $c_t^i = \left\{C(l) \,\middle|\,l \in \mathrm{Adj}_i\right\}$. 
\end{itemize}

\par Therefore, the local observation can be denoted as $o_t^i=(d_t^i,u_t^i,c_t^i)$.
These features provide information about recent traffic variability and local link conditions.
All three can be obtained through local measurements or configuration queries, without exchanging observations with other agents for each decision.

\subsubsection{Action Space $a_t^i$} Agent $i$ determines the traffic splitting ratios over the candidate paths from edge node $i$ to each destination edge node: $a_t^i = \left(g_p(t)\right)_{d \in \mathcal{N}^{1}\setminus\{i\},\,p \in P_{id}}$.
For each destination, the ratios satisfy the nonnegativity and unit-sum constraints in \eqref{eq2}.
As illustrated in Fig.~\ref{fig5}, the edge node translates the actor output into local updates to its data-plane traffic-to-path mapping table.

\subsubsection{Global Training State $s_t$} The global state comprises the local observations of all agents, i.e., $o_t=\{o_t^1,\ldots,o_t^N\}$, and the intermediate-node link states $k_t$ that are not observable by the edge agents.
Accordingly, the global state can be denoted as $s_t=(o_t,k_t)$.

\subsubsection{Global Reward $r_t$} MATE uses a sensitivity penalty as a soft incentive for the stability-dependent protection described in~\eqref{eq9}.
The reward combines realized MLU with variability-weighted path sensitivity:

\begin{equation}\label{eq12}
    r_t =
    -U_{\max}^{t} \cdot 
    \exp\!\left(\hat{A}_t\right),   
\end{equation}
where $U_{\max}^{t}$ is the MLU resulting from the joint action at time $t$, and $\hat{A}_t$ is the normalized regularization cost. 
The unnormalized cost can be defined as

\begin{equation}\label{eq13}
      A_t = \sum_{\substack{i,d \in \mathcal{N}^{1}\\i\neq d}} \mathrm{CV}_{id} \cdot \sum_{p \in P_{id}} S_p^{id}(t).
\end{equation}

\par For historical demand with positive mean $\mu_{id}$ and standard deviation $\sigma_{id}$, $\mathrm{CV}_{id}=\sigma_{id}/\mu_{id}$ is the coefficient of variation.
This dimensionless statistic measures variability relative to the mean, allowing comparison across demands with different magnitudes~\cite{nistcv}.
Using $S_p^{id}(t)=g_p(t)/C_p$, each path contributes $\mathrm{CV}_{id}g_p(t)/C_p$ to the regularization cost.
The coefficient associated with its splitting fraction is therefore $\mathrm{CV}_{id}/C_p$. 
Allocations to lower-capacity paths incur a greater cost, and this capacity preference is weighted more strongly for demands with greater historical variability. 
This design relates historical demand variation to the normalized load impact of additional traffic.

\par The two reward components have complementary roles.
The MLU term captures aggregate congestion and competition among demands on shared links. 
The regularizer introduces a variability-dependent preference among paths with different bottleneck capacities. 
For a demand whose candidate paths all have the same bottleneck capacity $C$, $\sum_{p\in P_{id}}g_p(t)/C_p=1/C$ for every feasible split.
In this case, congestion-aware splitting is determined by the MLU component and shared-link loads. 
With heterogeneous path capacities, the regularizer additionally favors allocations with a smaller variability-weighted capacity cost.

\par To control changes in the numerical scale of the penalty, the trainer normalizes $A_t$ using an exponential moving average (EMA)~\cite{nist_ema}:

\begin{equation}\label{eq14}
    \hat{A}_t = \frac{A_t}{\varepsilon + \operatorname{EMA}_t(A)},
\end{equation}
where $\operatorname{EMA}_t(A)$ can be expressed as 

\begin{equation}\label{eq15}
    \operatorname{EMA}_t(A) = \left(1-\zeta\right)\operatorname{EMA}_{t-1}(A) + \zeta A_t.
\end{equation}

\par The constant $\varepsilon>0$ prevents division by zero and is set to a small value (i.e., $10^{-8}$). 
The smoothing factor $\zeta \in (0, 1]$ controls the rate at which the EMA adapts to changes in $A_t$.
We use $\zeta = 1/w$ to relate the smoothing timescale to the traffic-history window.
For a fixed nonnegative preceding EMA value, the normalized cost increases monotonically with $A_t$. 
Consequently, at the same positive MLU and with the same normalization history, an allocation with a smaller variability-weighted capacity
cost receives a larger reward. 
With a nonnegative EMA initialization, the current-step update also gives $\operatorname{EMA}_t(A)\geq\zeta A_t$, yielding $0\leq\hat{A}_t\leq1/\zeta$.
The EMA rescales the regularization cost according to recent training history. Equation~\eqref{eq9} specifies explicit per-path sensitivity limits, while \eqref{eq12}--\eqref{eq15} provide a soft congestion-and-capacity objective for policy learning.

\subsubsection{Neural Network Architecture}

Each edge agent maintains an actor network $\pi_i(o_t^i\mid\theta_i)$, which maps its local observation $o_t^i$ to a traffic splitting action $a_t^i$.
A shared global critic $Q(s_t,a_t\mid\phi)$ takes the global state and joint action $a_t=(a_t^1,\ldots,a_t^N)$ as inputs and estimates the discounted return.
Its evaluations of the selected and counterfactual joint actions are used to derive agent-specific difference rewards for actor training.
Both the actors and the critic are fully connected neural networks.
Following the target-network construction used in DDPG and MADDPG~\cite{ddpg,maddpg}, each has a trainable behavior network and
a target network with the same architecture.
Actor $i$ has behavior parameters $\theta_i^b$ and target parameters $\theta_i^m$, and the corresponding critic parameters are
$\phi^b$ and $\phi^m$.

\subsubsection{Experience Collection and Traffic Replay}
\par Algorithm~\ref{alg1} summarizes the training procedure.
We first randomly initialize the behavior actors and critic, copy their parameters to the corresponding target networks, and initialize a finite replay buffer $RB$~\cite{ddpg}.
At the beginning of each episode, the default routing protocol initializes traffic allocation to obtain the initial network state.
At each time step $t=0,\ldots,T$, each agent $i$ selects $a_t^i=\pi_i(o_t^i\mid\theta_i^b)$ from its local observation.
The virtual network environment executes the resulting joint action $a_t$.
Then, the trainer computes the reward using \eqref{eq12} and stores $(s_t,a_t,r_t,s_{t+1},\chi_t)$ in $RB$, where $\chi_t=1$ denotes termination after the transition and $\chi_t=0$ otherwise.
The local observations required by the actors are components of $s_t$ and $s_{t+1}$.
Once $|RB|\geq V$, training updates use minibatches of $V$ randomly sampled transitions.


\par To revisit traffic conditions while retaining their temporal variation, we adopt short-sequence replay inspired by RedTE~\cite{redte}.
Within each training epoch, several contiguous subsequences are selected from the historical TM series and replayed repeatedly.
Each rollout preserves the chronological order within its subsequence, with traffic-history windows constructed accordingly.
Minibatches for parameter updates are still sampled randomly from $RB$.
This distinguishes chronological traffic replay from random experience sampling and provides repeated exposure to each selected traffic regime.


\subsubsection{Counterfactual Credit Assignment and Actor Updates} \label{sec:counterfactual}
During centralized training, MATE uses the shared global critic to derive agent-specific credit signals. 
Following counterfactual credit assignment~\cite{coma,guodistri}, the difference reward for agent $i$ is defined as

\begin{equation}\label{eq16}
    R_t^i = Q(s_t,a_t\mid\phi^b) - \frac{1}{H}\sum_{h=1}^{H} Q(s_t,a_t'^{h,i}\mid\phi^b).
\end{equation}
where $H\geq 1$ is the number of alternative actions sampled for agent $i$ at each evaluated state. 
Each alternative $\bar{a}_t^{h,i}$ contains the splitting ratios from node $i$ to all destination edge nodes.
For each destination $d\in\mathcal{N}^{1}\setminus\{i\}$, its components satisfy
\begin{equation*}
    0\leq \bar{g}_{t,p}^{h,i}\leq 1, \qquad
    \sum_{p\in P_{id}}\bar{g}_{t,p}^{h,i}=1.
\end{equation*}

\par Let $\mathcal{A}_i$ denote the resulting splitting-action space.
The counterfactual joint action $a_t'^{h,i}$ is obtained by replacing the $i$th component of $a_t$ with $\bar{a}_t^{h,i}\in\mathcal{A}_i$, while keeping $s_t$ and all other agents' actions fixed.
The average of the $H$ counterfactual evaluations forms the baseline for agent $i$. 
A positive $R_t^i$ indicates that the selected action has a higher estimated global return than the sampled alternatives on average.

\par These difference rewards provide agent-specific credit for actor training.
The critic continues to learn from the shared global reward $r_t$, so both joint-action evaluation and credit assignment are grounded in the same network-wide objective.

\par For each sampled state $s_v$, let $\widehat{a}_v^{\,i}$ be the joint action formed from the current actors. 
During the update of actor $i$, its components are

\begin{equation*}
    \bigl(\widehat{a}_v^{\,i}\bigr)_j =
    \begin{cases}
        \pi_i(o_v^i\mid\theta_i^b), & j=i,\\
        \operatorname{sg}\!\left(
        \pi_j(o_v^j\mid\theta_j^b)\right), & j\neq i,
    \end{cases}
\end{equation*}
where $\operatorname{sg}(\cdot)$ denotes
the stop-gradient operation.

\par The corresponding difference reward $\widehat{R}_v^i$ is computed using~\eqref{eq16}, with $\widehat{a}_v^{\,i}$ as the evaluated joint action.
The actor is trained using the credit-weighted surrogate

\begin{equation}\label{eq17}
    L_{\theta_i^b} =
    -\frac{1}{V}\sum_{v=1}^{V} \operatorname{sg}\!\left(\widehat{R}_v^i\right) Q(s_v,\widehat{a}_v^{\,i}\mid\phi^b),
\end{equation}
where $V$ is the minibatch size and $v$ indexes the sampled transitions.

\par The entire credit weight $\widehat{R}_v^i$ is treated as constant during differentiation. 
The critic parameters and the other actors' outputs are also held fixed.
Gradients are retained through the critic's action input to actor $i$, so the credit weight scales the resulting actor gradient.

\begin{algorithm}[h]
\begin{spacing}{1.0}
    \caption{Offline Centralized Training of MATE}
    \label{alg1}

    \KwIn{Topology $G=(\mathcal{N},\mathcal{L})$,
    historical TMs, and counterfactual sample count $H$.}
    \KwOut{Trained actor parameters
    $\{\theta_i^b\}_{i=1}^{N}$.}

    Randomly initialize actor parameters
    $\{\theta_i^b\}_{i=1}^{N}$
    and critic parameters $\phi^b$\;

    Set $\theta_i^m\leftarrow\theta_i^b$, $\forall i$,
    $\phi^m\leftarrow\phi^b$, and $RB\leftarrow\emptyset$\;

    \For{$e=1,\ldots,E$}{
        Initialize $s_0$ using default routing\;

        \For{$t=0,\ldots,T$}{
            Select $a_t^i=\pi_i(o_t^i\mid\theta_i^b)$,
            $\forall i$, and execute
            $a_t=(a_t^1,\ldots,a_t^N)$\;

            Observe and store
            $(s_t,a_t,r_t,s_{t+1},
            \{o_{t+1}^i\}_{i=1}^{N},\chi_t)$
            in $RB$\;

            \If{$|RB|\geq V$}{
                Sample a minibatch of $V$ transitions
                from $RB$\;

                Form $\widehat{a}_v^{\,i}$ from current
                actors, detaching outputs of actors
                $j\neq i$, $\forall i,v$\;

                Compute $\widehat{R}_v^i$
                using \eqref{eq16} with $H$ alternatives
                from $\mathcal{A}_i$, varying only
                agent $i$'s action\;

                Compute $L_{\theta_i^b}$
                using \eqref{eq17} with detached
                credit weights and gradients through
                actor $i$\;

                Compute TD targets $y_v$
                and critic loss $L_{\phi^b}$
                using \eqref{eq18} and \eqref{eq19}\;

                $\theta_i^b\leftarrow\theta_i^b
                -\eta_{\pi}
                \nabla_{\theta_i^b}L_{\theta_i^b}$,
                $\forall i$\;

                $\theta_i^m\leftarrow\tau\theta_i^b
                +(1-\tau)\theta_i^m$,
                $\forall i$\;

                $\phi^b\leftarrow\phi^b
                -\eta_Q\nabla_{\phi^b}L_{\phi^b}$\;

                $\phi^m\leftarrow\tau\phi^b
                +(1-\tau)\phi^m$\;
            }
        }
    }

    \Return{$\{\theta_i^b\}_{i=1}^{N}$}\;
\end{spacing}
\end{algorithm}

\subsubsection{Critic and Target-Network Updates}
The shared critic learns the return associated with the global reward $r_t$ in~\eqref{eq12}. 
Its action-value estimates also provide the agent-specific credit signals in~\eqref{eq16} for actor learning. 
For each sampled transition, the temporal difference target is

\begin{equation}\label{eq18}
    y_t = r_t + \gamma \cdot \left(1-\chi_t\right) \cdot Q\left(s_{t+1},a_{t+1}^{\prime} \mid \phi^m\right),
\end{equation}
where $\gamma \in[0,1)$ denotes the discount factor, and $\chi_t$ equals one if the episode terminates after the transition and zero otherwise. 
The next joint action $a_{t+1}^{\prime}$ can be generated by the target actors using their respective next local observations, i.e., $a_{t+1}^{\prime} = \bigl(\pi_j(o_{t+1}^j \mid \theta_j^m)\bigr)_{j=1}^{N}$.

\par The target is treated as constant when updating the behavior critic. The critic minimizes the minibatch mean squared error~\cite{ddpg,maddpg}:

\begin{equation}\label{eq19}
    L_{\phi^b} = \frac{1}{V} \sum_{v=1}^{V} \left(Q(s_v,a_v\mid\phi^b) - y_v\right)^2.
\end{equation}

\par The target networks track the behavior networks through soft updates~\cite{ddpg}:

\begin{subequations}\label{eq20}
    \begin{align}
        & \theta_i^m \leftarrow \tau \theta_i^b + \left(1-\tau\right) \theta_i^m, \\
        & \phi^m \leftarrow \tau \phi^b + \left(1-\tau\right)\phi^m,
    \end{align}
\end{subequations}
where $\tau \in(0,1]$ controls the rate at which the target parameters track the behavior parameters. After offline training, the controller distributes each learned actor to its corresponding edge router for online traffic scheduling.

\vspace{-1em}

\subsection{Online Execution Phase}

\par During online execution, each edge node independently determines traffic-splitting ratios without inter-agent communication.
Its actor network takes local observations as inputs, including historical traffic demands to other edge nodes and local link states.
The resulting splitting ratios specify traffic allocation among the preconfigured candidate paths. Ingress nodes implement this allocation using the SRHs associated with the corresponding segment lists~\cite{rfc9256,SRv6}.
By exploiting recent traffic history and the stability-dependent sensitivity penalties introduced during training, MATE mitigates the impact of traffic bursts while limiting the efficiency cost of conservative scheduling.

\par MATE performs computationally intensive training offline.
Online operation involves only neural network forward inference, candidate-path selection, and lightweight traffic-allocation computations, without backpropagation or local gradient updates.
This avoids repeated centralized optimization during online operation.
SRv6 encapsulation and forwarding use the existing data-plane capabilities of network devices.
The agents operate in the control plane or edge control modules and do not participate in per-packet forwarding.
When a network failure occurs, edge nodes receive notifications from the centralized controller or the underlying routing protocol. 
Each node identifies the affected candidate paths and assigns a large value, such as $1000\%$, to the relevant link-utilization inputs to discourage traffic allocation to these paths~\cite{redte}.
The controller continues to collect training data and periodically retrains the models offline, for example, once a week, to mitigate performance degradation caused by traffic distribution shifts~\cite{redte}.
The updated actor models are then distributed to the corresponding edge routers via gRPC~\cite{grpc}.

\section{Performance Evaluation}

\par We evaluate MATE on a geographically distributed WAN testbed built on CENI and through simulations on the Sanet topology from TopologyZoo~\cite{topozoo}, with 43 nodes and 90 links.

\begin{figure}[h]
\centering
\includegraphics[scale=0.16]{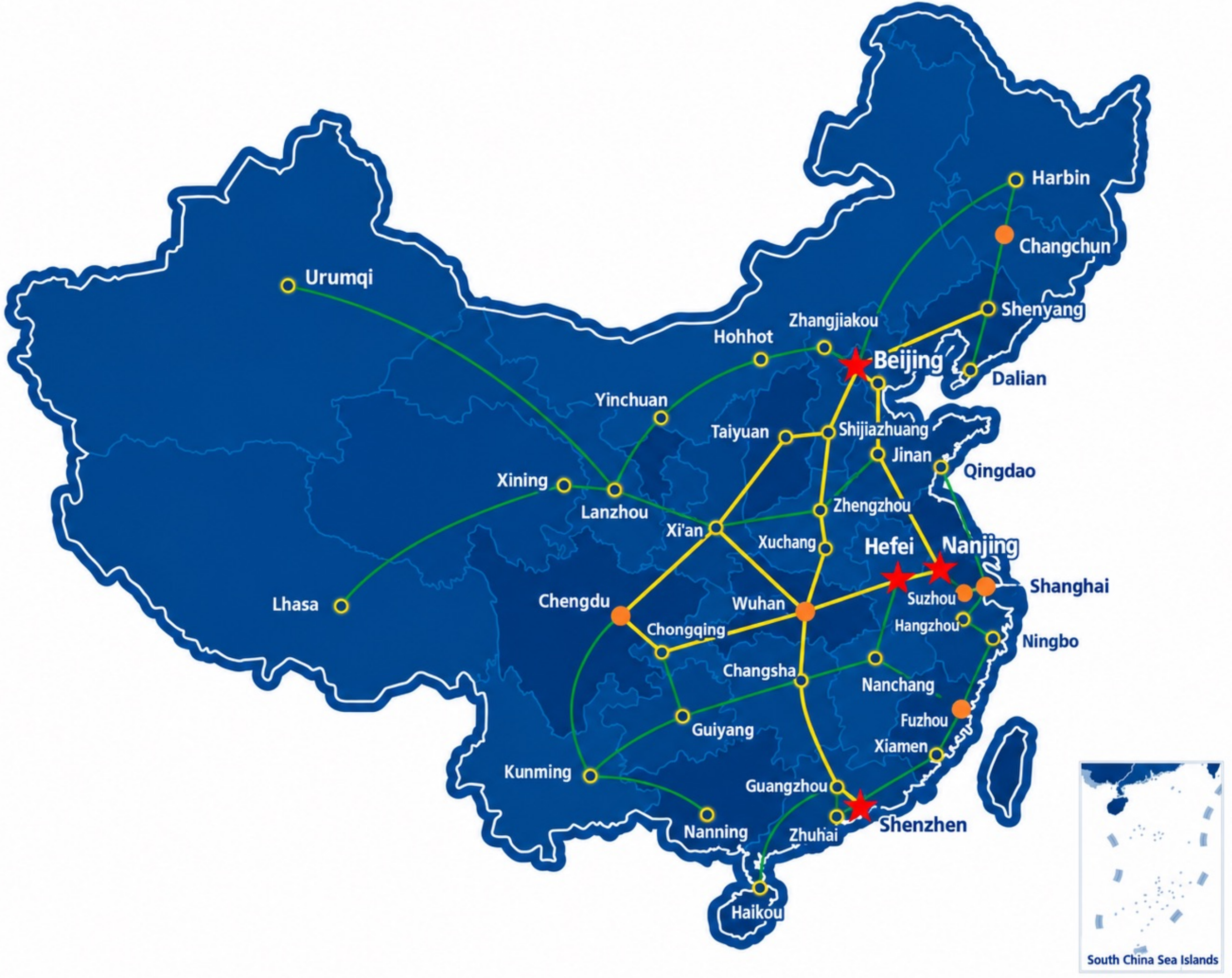}
\caption{\small China’s Future Network Experimental Infrastructure—CENI-WAN Testbed \cite{ceni}.}
\vspace{-1.5em}
\label{fig7}
\end{figure}

\subsection{CENI-WAN Testbed and Evaluation Setup}

\par
\begin{figure}[h]
    \centering
    \begin{subfigure}[b]{0.7\linewidth}
        \centering
        \includegraphics[width=\linewidth]{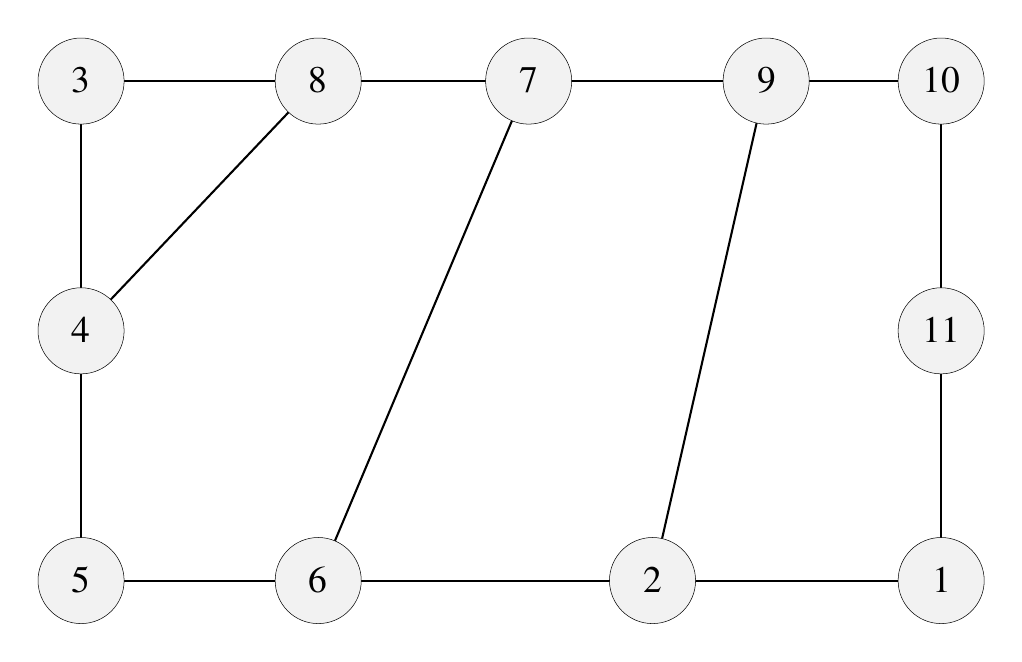}
        \caption{CENI-Abilene}
    \end{subfigure}

    \par\medskip

    \begin{subfigure}[b]{0.7\linewidth}
        \centering
        \includegraphics[width=\linewidth]{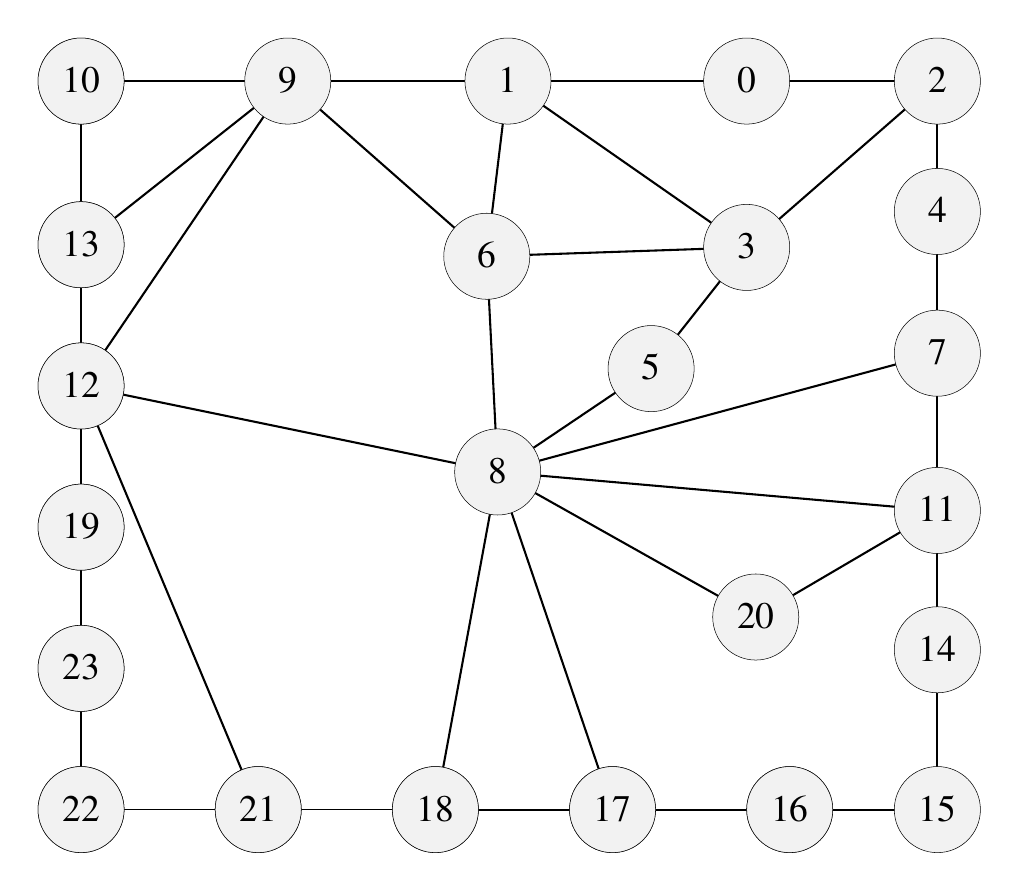}
        \caption{CENI-G\'EANT}
    \end{subfigure}
    \caption{Network topologies of CENI-Abilene and CENI-G\'EANT.}
    \label{fig8}
    \vspace{-1em}
\end{figure}

 We conduct experiments on the CENI-WAN testbed, as illustrated in Fig. \ref{fig7}. 
To evaluate proactive scheduling under realistic temporal traffic variations, we use measured traffic traces from operational networks.
Using CENI's network-slicing capabilities, we instantiate testbed networks isomorphic to G\'EANT~\cite{geant} and the Abilene topology from TopologyZoo~\cite{topozoo}, referred to as CENI-G\'EANT and CENI-Abilene, respectively.
The corresponding public TM traces have sampling intervals of 15 minutes for G\'EANT~\cite{geant} and 5 minutes for Abilene~\cite{AbileneTM}. Fig.~\ref{fig8} shows the two topologies.

\par To broaden the evaluation, we also construct CENI-Bandcon using the Bandcon topology from TopologyZoo~\cite{topozoo}.
Since measured TMs are unavailable for our Bandcon experiments, we generate 2048 synthetic TMs using a gravity model~\cite{gravitymodel}.
Of these TMs, $75\%$ are used for offline training and the remaining $25\%$ for online scheduling evaluation.
Traffic demand magnitudes are scaled to accommodate the experimental environment. 
In Table~\ref{toponum2}, we summarize the numbers of nodes and links in the three testbed topologies.
For topologies without explicit edge and intermediate node designations, we designate the $30\%$--$35\%$ of nodes with the highest degrees as intermediate nodes and the remaining nodes as edge nodes.
The testbed networks are instantiated using VLAN~\cite{ieee8021q}, VXLAN~\cite{rfc7348}, and segment routing~\cite{rfc8402,SRv6}.

\begin{table}[h]
    \centering
    \caption{Evaluation Topologies}
    \label{toponum2}
    \renewcommand{\arraystretch}{1.1}
    \begin{tabular}{@{}lcc@{}}
        \toprule
        Topology & Nodes & Links \\
        \midrule
        CENI-Abilene & 11 & 28 \\
        CENI-G\'EANT & 24 & 74 \\
        CENI-Bandcon & 24 & 74 \\
        \bottomrule
    \end{tabular}
\end{table}

\par Edge nodes are distributed across cities outside Hefei, including Suzhou, Beijing, and Shanghai, to capture practical communication delays and provide geographical and network-domain separation.
All edge nodes are SRv6-enabled and generate proactive splitting decisions from local link states and historical traffic measurements, enforcing them through SRv6 source encapsulation~\cite{SRv6}.
Intermediate nodes are hosted at research institutes and universities across China, with SRv6 enabled on a subset. 
We evaluate different SRv6 deployment ratios using candidate paths realizable under the corresponding enabled-node configuration and underlying IP routing.
Network nodes and the controller each have four CPU cores and run Ubuntu~18.04.
Each network node has 12~GB of memory. 
The centralized controller, hosted at the Institute of Advanced Technology, University of Science and Technology of China, has an RTX~3080 GPU and 200~GB of storage.
It handles candidate-path computation and updates, SRv6 tunnel establishment, failure detection, topology and resource maintenance, and agent-model training and distribution.
OSPF provides underlying connectivity and routing reachability, while iPerf~\cite{iperf} generates traffic at rates specified by the corresponding TMs.

\par Table~\ref{para2} lists the offline training settings.
The counterfactual action constraints and actor-gradient handling are specified in Sec.~\ref{sec:counterfactual}.
The reported training curve converges under the evaluated configuration, while Sec.~\ref{相关因素} separately examines
the tested actor and critic architectures.

\begin{table}[h]
    \centering
    \caption{MATE Training Parameters}
    \label{para2}
    \footnotesize
    \renewcommand{\arraystretch}{1.15}
    \begin{tabular}{@{}l@{\hspace{6pt}}c@{}}
        \toprule
        Parameter & Value \\
        \midrule
        Actor layers and widths
        & $3$; $(128,64,64)$ \\

        Critic layers and widths
        & $3$; $(128,64,128)$ \\

        Optimizer
        & Adam \\

        Actor and critic learning rates
        & $10^{-4}$, $2\times10^{-4}$ \\

        Replay buffer capacity
        & $9600$ \\

        Minibatch size $V$
        & $64$ \\

        Discount factor $\gamma$
        & $0.96$ \\

        Training episodes $E$
        & $1000$ \\

        Target-network soft-update coefficient $\tau$
        & $0.005$ \\

        Traffic history window $w$
        & $15$ \\

        Reward normalization parameters $(\varepsilon,\zeta)$
        & $(10^{-8},0.067)$ \\
        \bottomrule
    \end{tabular}
\end{table}

\par We compare MATE with the following reference implementations
to evaluate scheduling performance under the tested conditions.

\begin{itemize}
    \item \textbf{TEAL}~\cite{teal}: Using global connectivity, link capacities, and demands, TEAL combines a flow-centric GNN with multi-agent DRL and a few ADMM iterations to refine allocations.
    Since instantaneous demands are unavailable in our evaluation, we use the latest completed interval's TM to configure the following interval.

    \item \textbf{DOTE}~\cite{dote}: This centralized learning-based method directly maps historical TMs to the next interval's TE configuration through a global decision model.

    \item \textbf{FIGRET}~\cite{figret}: This centralized proactive method maps historical TMs to routing configurations. 
    Its loss weights path-sensitivity penalties by demand variance to protect against bursts. 
    Topology and capacities enter the training objective, but current topology and capacity features are not explicit inputs to its neural policy during inference.

    \item \textbf{COPE}~\cite{cope}: COPE optimizes expected-demand performance subject to worst-case guarantees against demand uncertainty, with substantial computational overhead~\cite{figret}.

    \item \textbf{MARL}: This distributed multi-agent ablation baseline removes the counterfactual credit weights from the actor objective.
    Each actor directly maximizes the shared critic value, equivalently using unit weights in~\eqref{eq17}:
    \begin{equation*}
      L_{\theta_i^b}^{\mathrm{MARL}} =
    -\frac{1}{V}\sum_{v=1}^{V}
    Q(s_v,\widehat{a}_v^{\,i}\mid\phi^b).
    \end{equation*}

    The critic parameters and the other actors' outputs are held fixed during each actor update. This objective does not require counterfactual sampling.

    \item \textbf{LP}: This oracle baseline uses Gurobi~\cite{gurobi} to solve $\mathcal{P}1$ for each realized test TM under the evaluated topology, link capacities, SRv6 deployment, and candidate-path set.
    The sensitivity constraint in~\eqref{eq9} is excluded.
    Its optimum serves as the normalization reference; instead of plotting the LP separately, we report each method's normalized MLU as follows:
    \begin{equation*}
      \text{Normalized MLU} = \frac{U_{\max}^{\mathrm{method}}}
{U_{\max}^{\mathrm{LP}}}.
    \end{equation*}
\end{itemize}

\begin{figure}[h]
    \centering
    \includegraphics[width=0.9\linewidth]
    {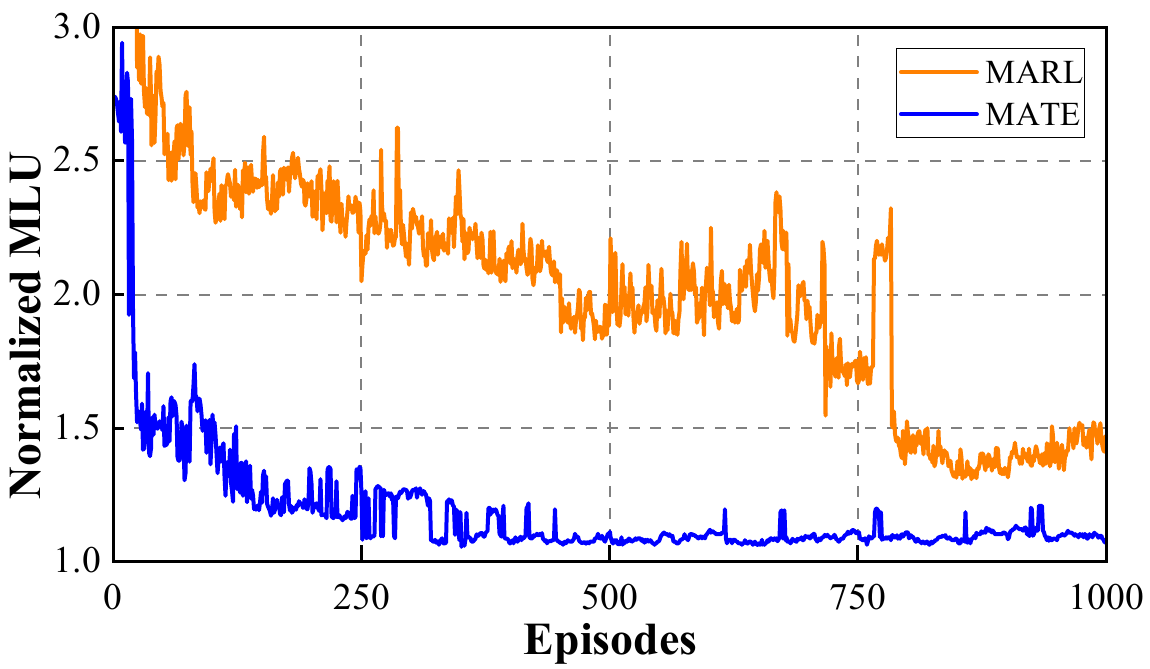}
    \caption{\small Training convergence of MATE and MARL on CENI-Abilene.}
    \label{mate收敛图}
    \vspace{-1em}
\end{figure}

\subsection{Performance on the CENI-WAN Testbed}\label{2真实}

\par We evaluate MATE on CENI-Abilene and CENI-G\'EANT using measured traffic time series, and on CENI-Bandcon using synthetic TMs.
These networks span different sizes, link structures, and traffic patterns.
We examine offline training convergence, normalized MLU distributions across SRv6 deployment ratios to assess proactive burst handling, and performance differences between measured and synthetic traffic.

\subsubsection{Training Convergence}

\par As shown in Fig.~\ref{mate收敛图}, both MATE and MARL exhibit fluctuations during early policy updates, followed by a decrease in normalized MLU.
Since the joint splitting decisions determine network-wide link loads, changes in the agents' policies can produce variations in the observed MLU during training.
MATE largely converges after approximately 500 episodes, whereas MARL reaches a relatively stable level after approximately 800 episodes.
MATE reaches approximately $1.1$, whereas MARL reaches $1.4$ after 1000 episodes, a reduction of approximately $21\%$.
MATE uses counterfactual comparisons under the shared global critic to derive agent-specific difference rewards for actor training~\cite{coma}.
MARL learns from a shared signal without this credit decomposition~\cite{coma}.
Under the stated ablation design, the faster convergence and lower normalized MLU are consistent with the benefits of agent-specific credit assignment in the evaluated
training configuration.

\begin{figure*}[tb]
    \centering
    \begin{subfigure}[b]{0.31\linewidth}
        \centering
        \includegraphics[width=\linewidth]{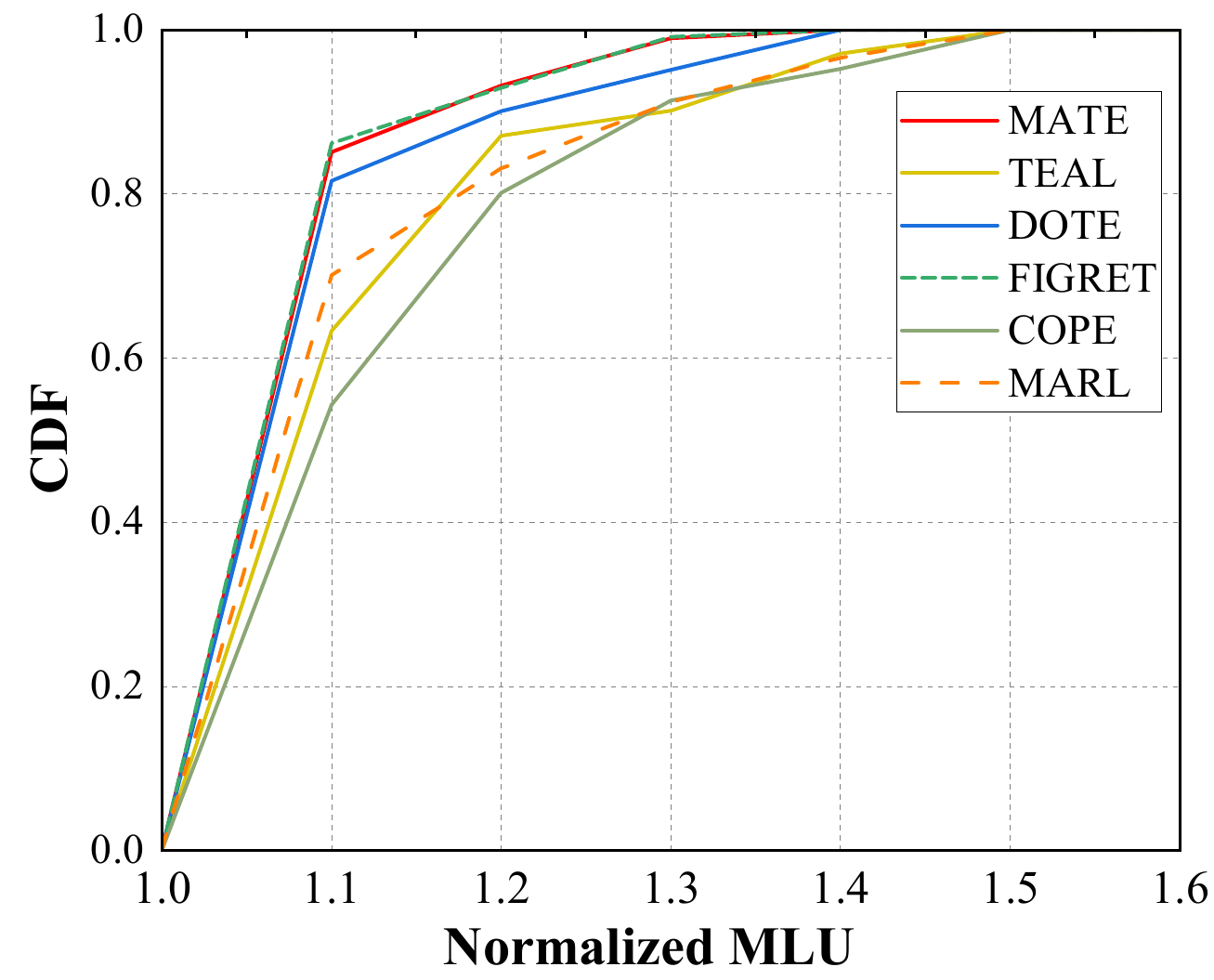}
        \caption{Deployment ratio: $0.3$}
    \end{subfigure}
    \hfill
    \begin{subfigure}[b]{0.31\linewidth}
        \centering
        \includegraphics[width=\linewidth]{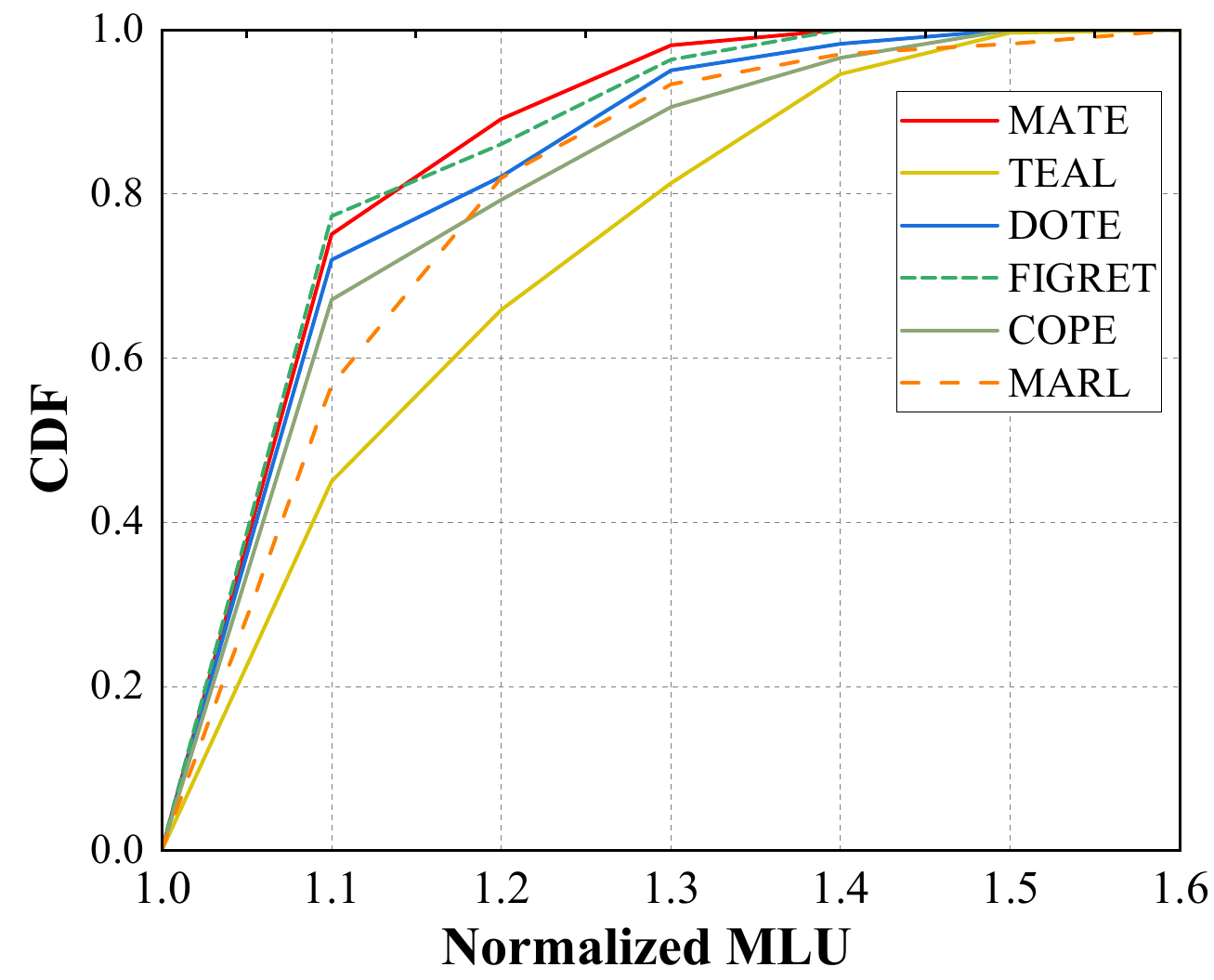}
        \caption{Deployment ratio: $0.6$}
    \end{subfigure}
    \hfill
    \begin{subfigure}[b]{0.31\linewidth}
        \centering
        \includegraphics[width=\linewidth]{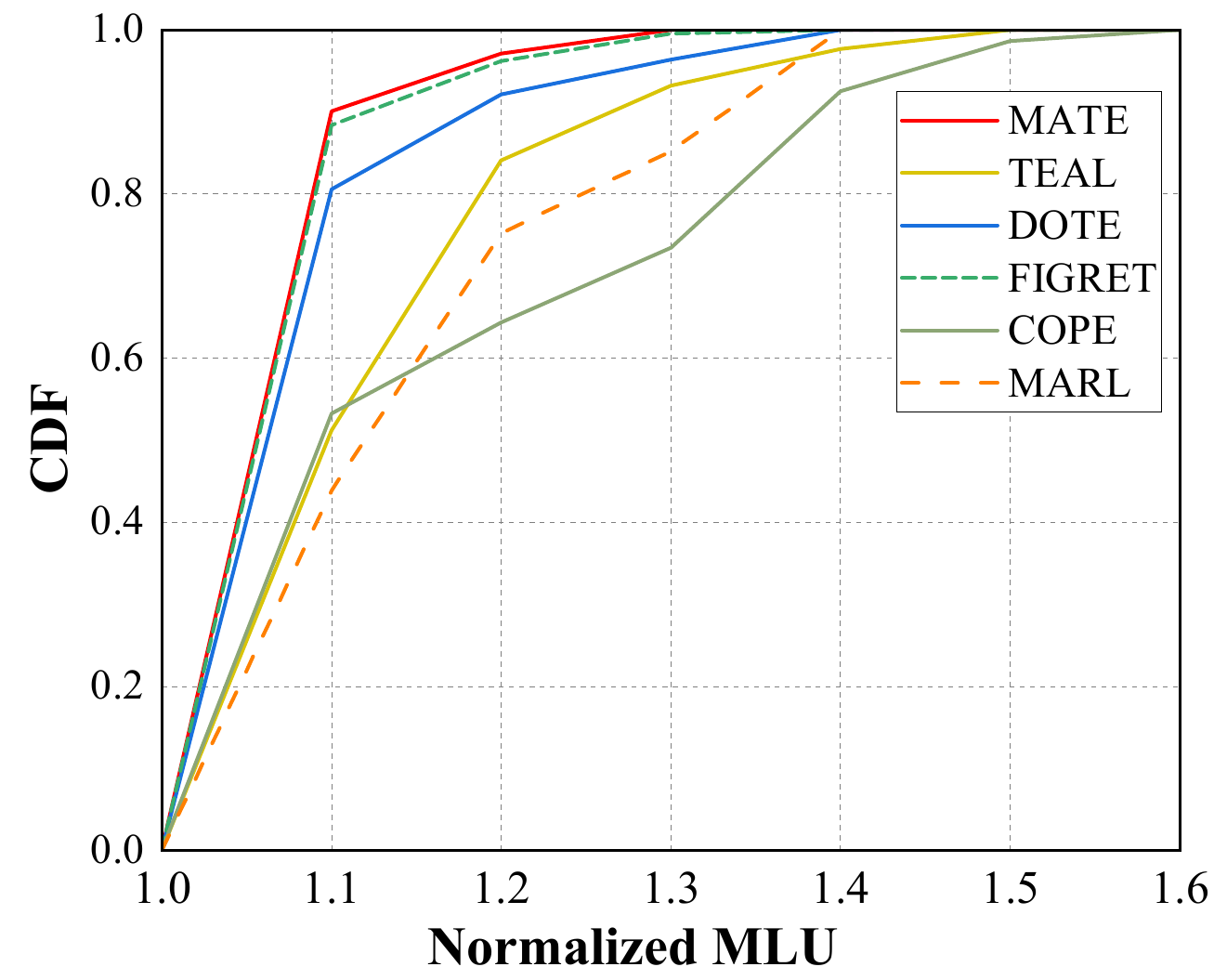}
        \caption{Deployment ratio: $1.0$}
    \end{subfigure}
    \caption{\small CDFs of normalized MLU on CENI-Abilene under different SRv6 deployment ratios.}
    \vspace{-0.5em}
    \label{Abilene前瞻性}
\end{figure*}

\begin{figure*}[tb]
    \centering
    \begin{subfigure}[b]{0.31\linewidth}
        \centering
    \includegraphics[width=\linewidth]
        {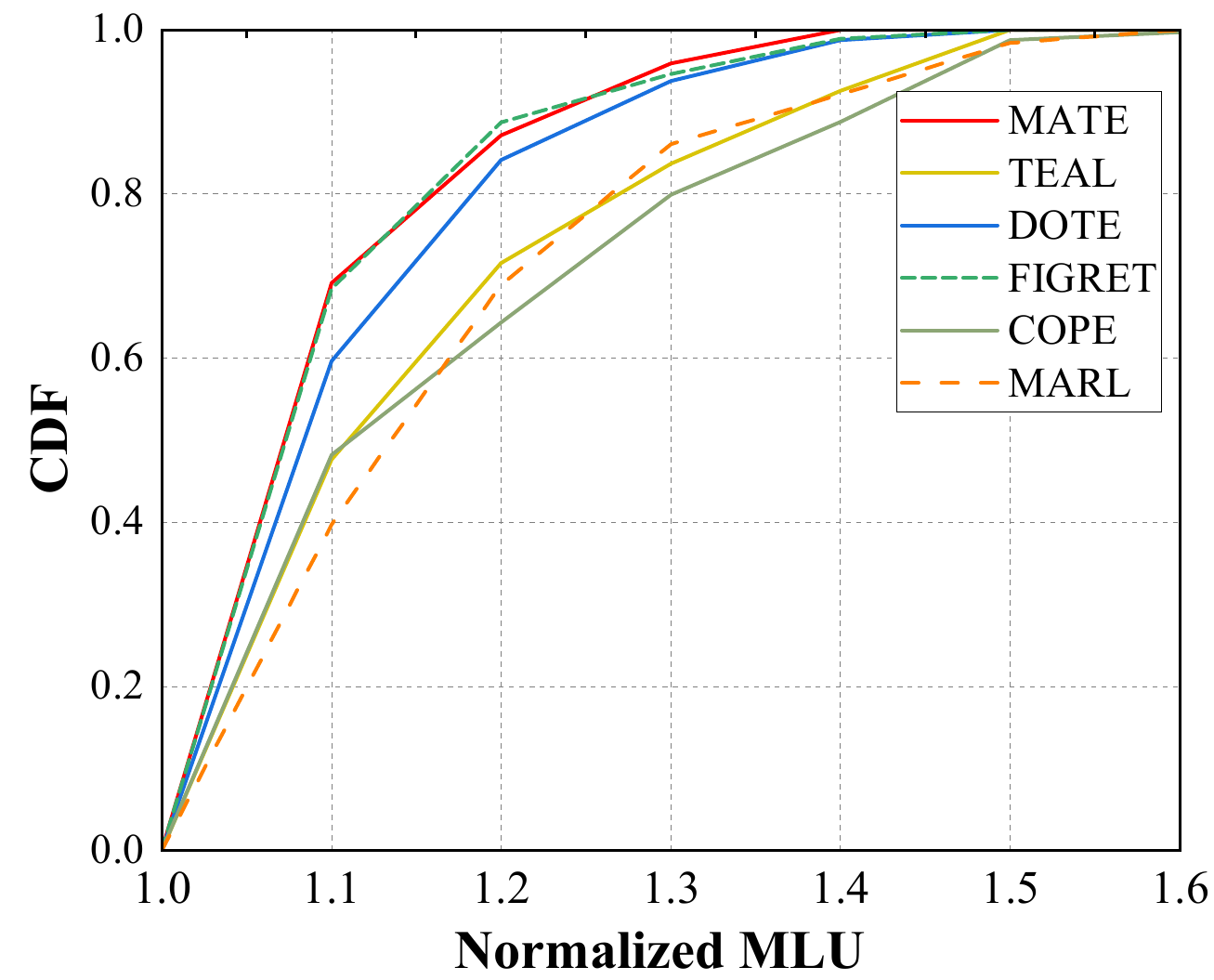}
        \caption{Deployment ratio: $0.3$}
    \end{subfigure}
    \hfill
    \begin{subfigure}[b]{0.31\linewidth}
        \centering
    \includegraphics[width=\linewidth]
        {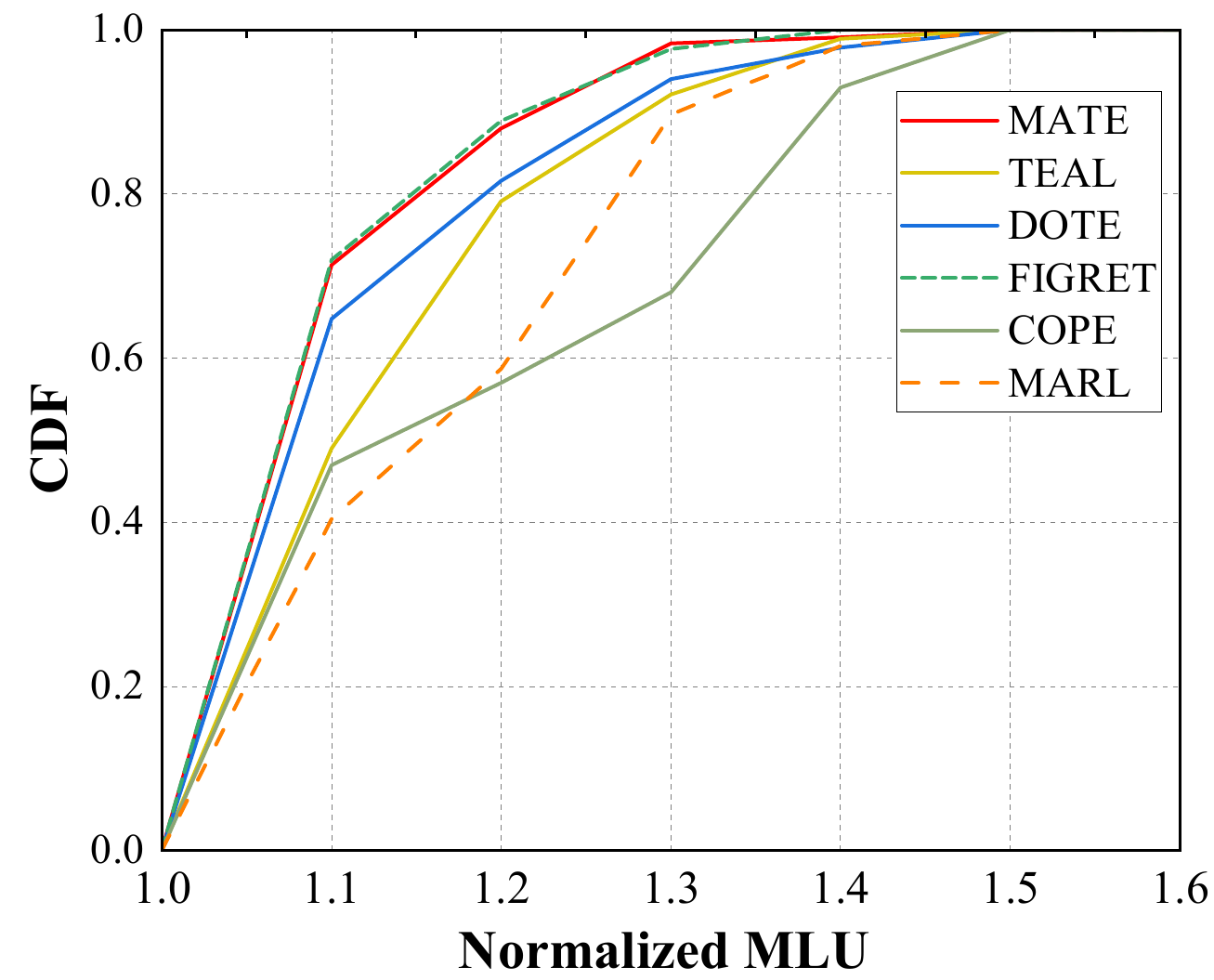}
        \caption{Deployment ratio: $0.6$}
    \end{subfigure}
    \hfill
    \begin{subfigure}[b]{0.31\linewidth}
        \centering
    \includegraphics[width=\linewidth]
        {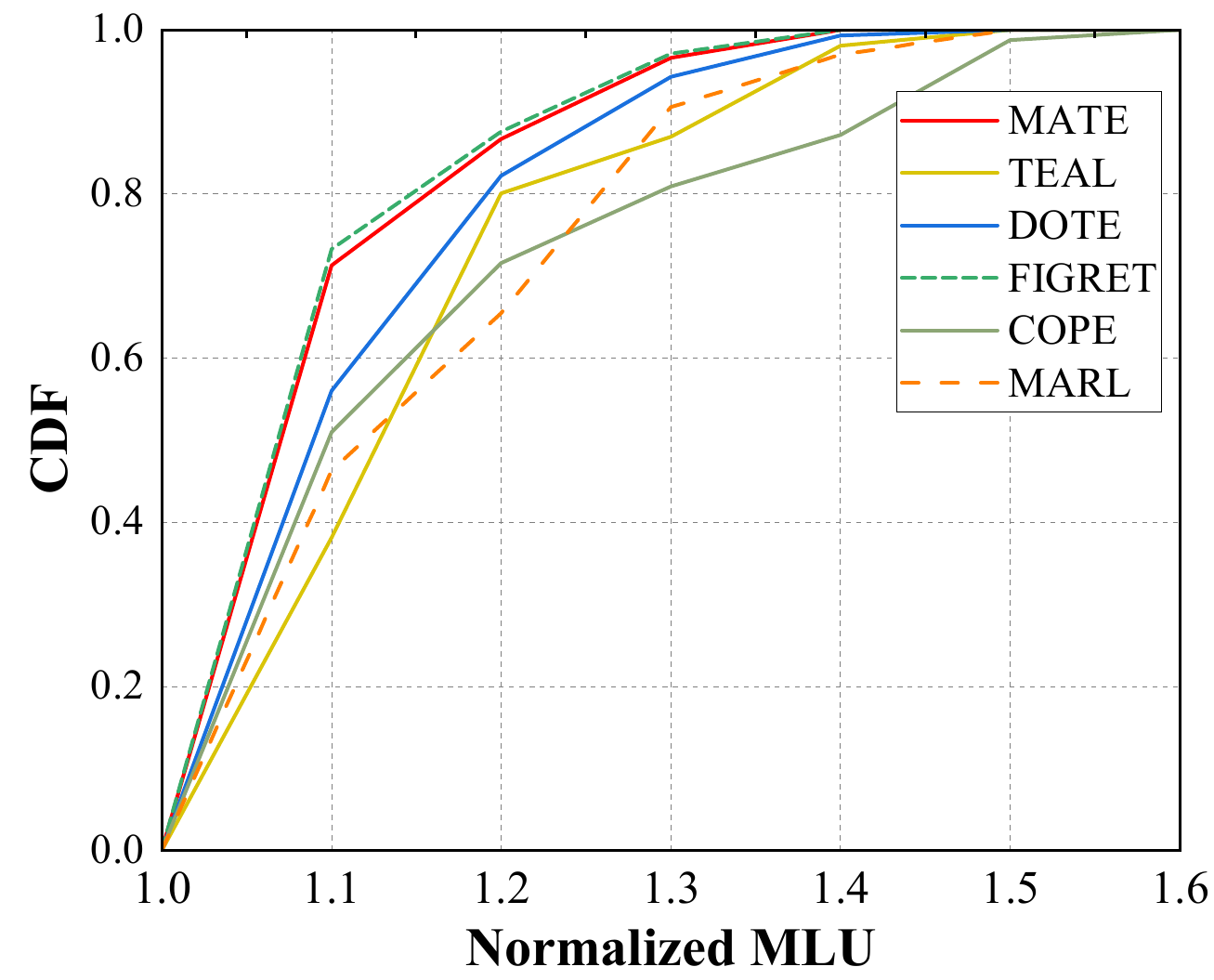}
        \caption{Deployment ratio: $1.0$}
    \end{subfigure}
    \caption{\small CDFs of normalized MLU on CENI-G\'EANT under different SRv6 deployment ratios.}
    \label{geant前瞻性}
    \vspace{-1em}
\end{figure*}

\subsubsection{Proactive Scheduling Performance}

\par
\begin{figure*}[tb]
    \centering
    \begin{subfigure}[b]{0.31\linewidth}
        \centering
        \includegraphics[width=\linewidth]
        {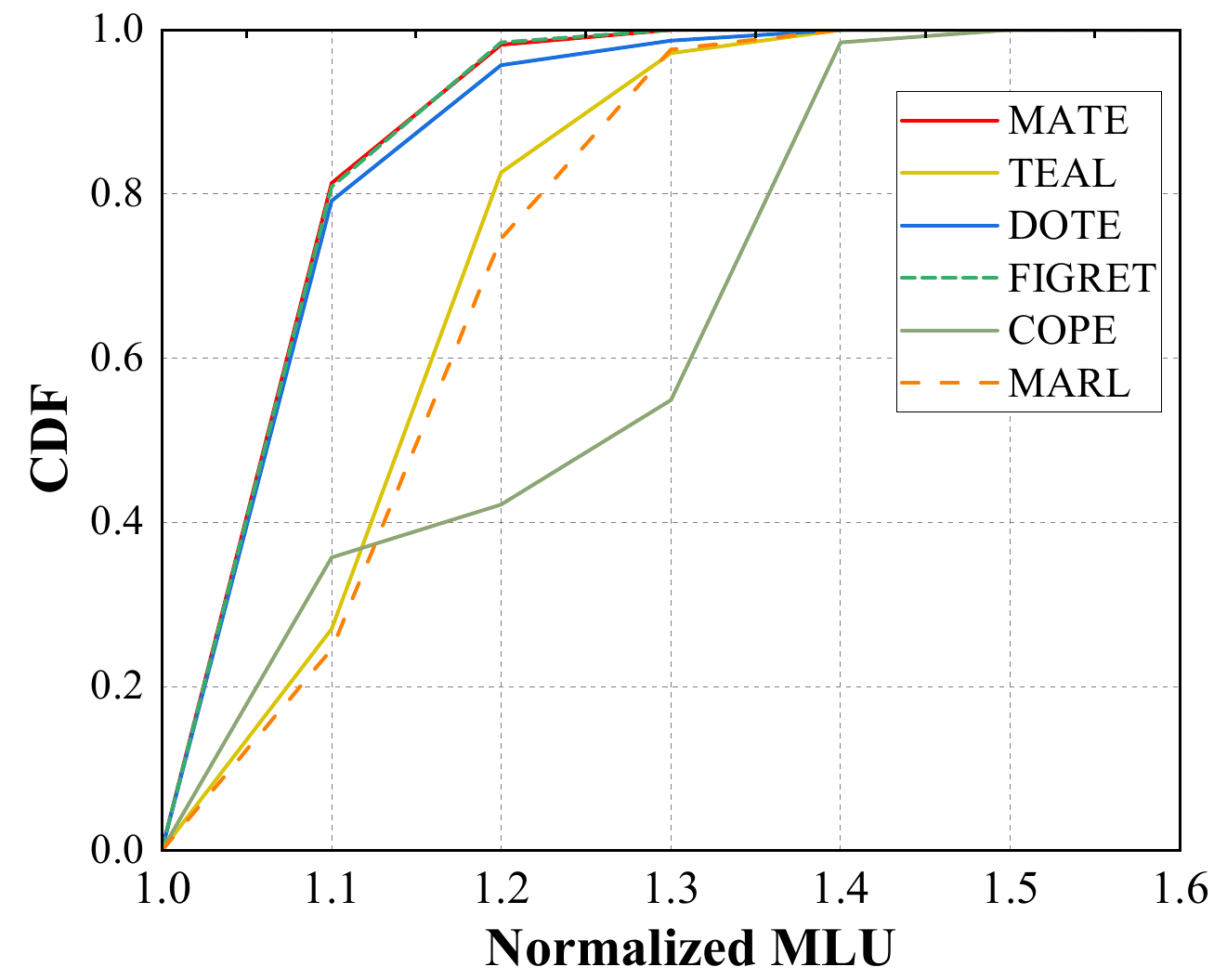}
        \caption{\small Deployment ratio: $0.3$}
    \end{subfigure}
    \hfill
    \begin{subfigure}[b]{0.31\linewidth}
        \centering
        \includegraphics[width=\linewidth]
        {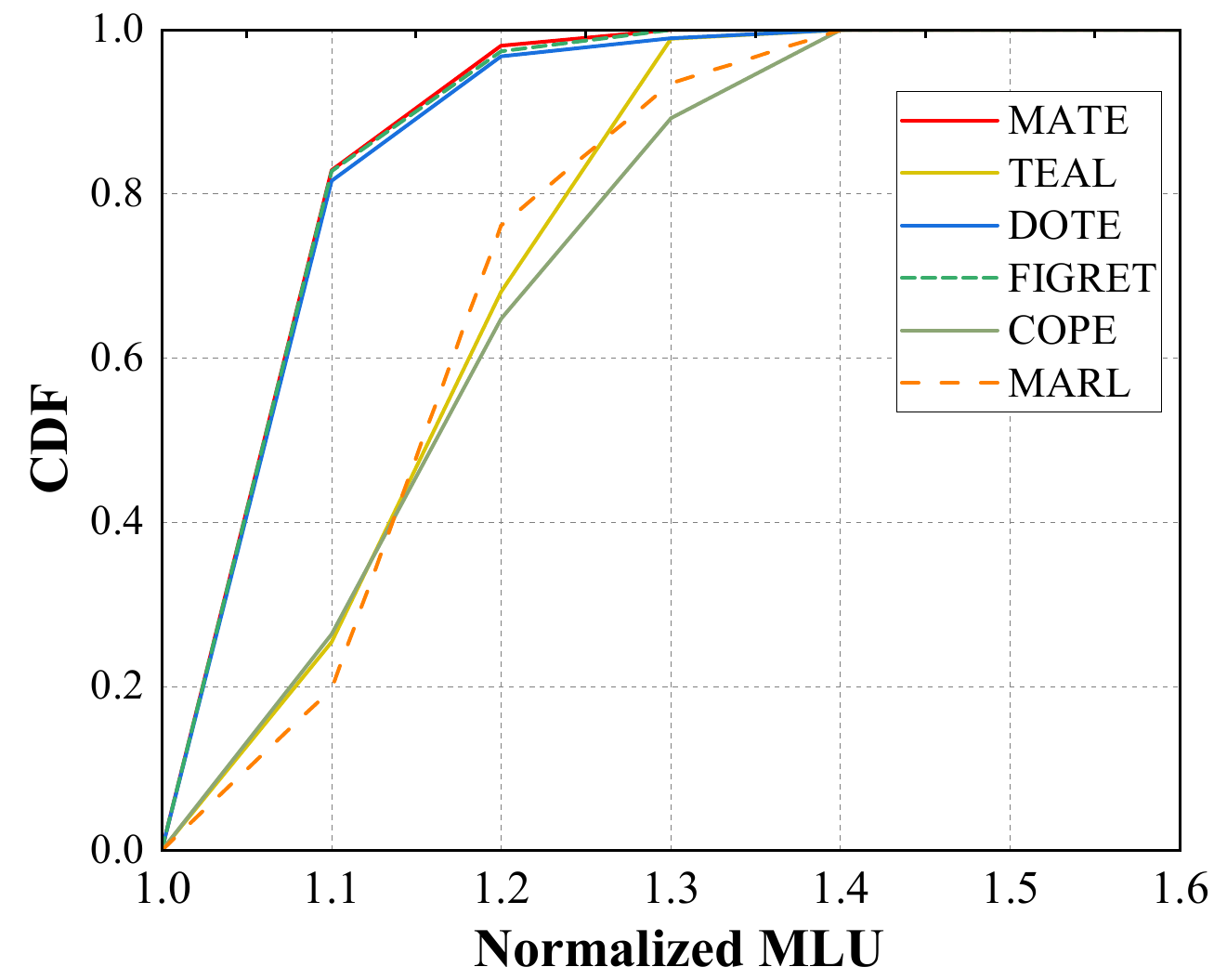}
        \caption{\small Deployment ratio: $0.6$}
    \end{subfigure}
    \hfill
    \begin{subfigure}[b]{0.31\linewidth}
        \centering
        \includegraphics[width=\linewidth]
        {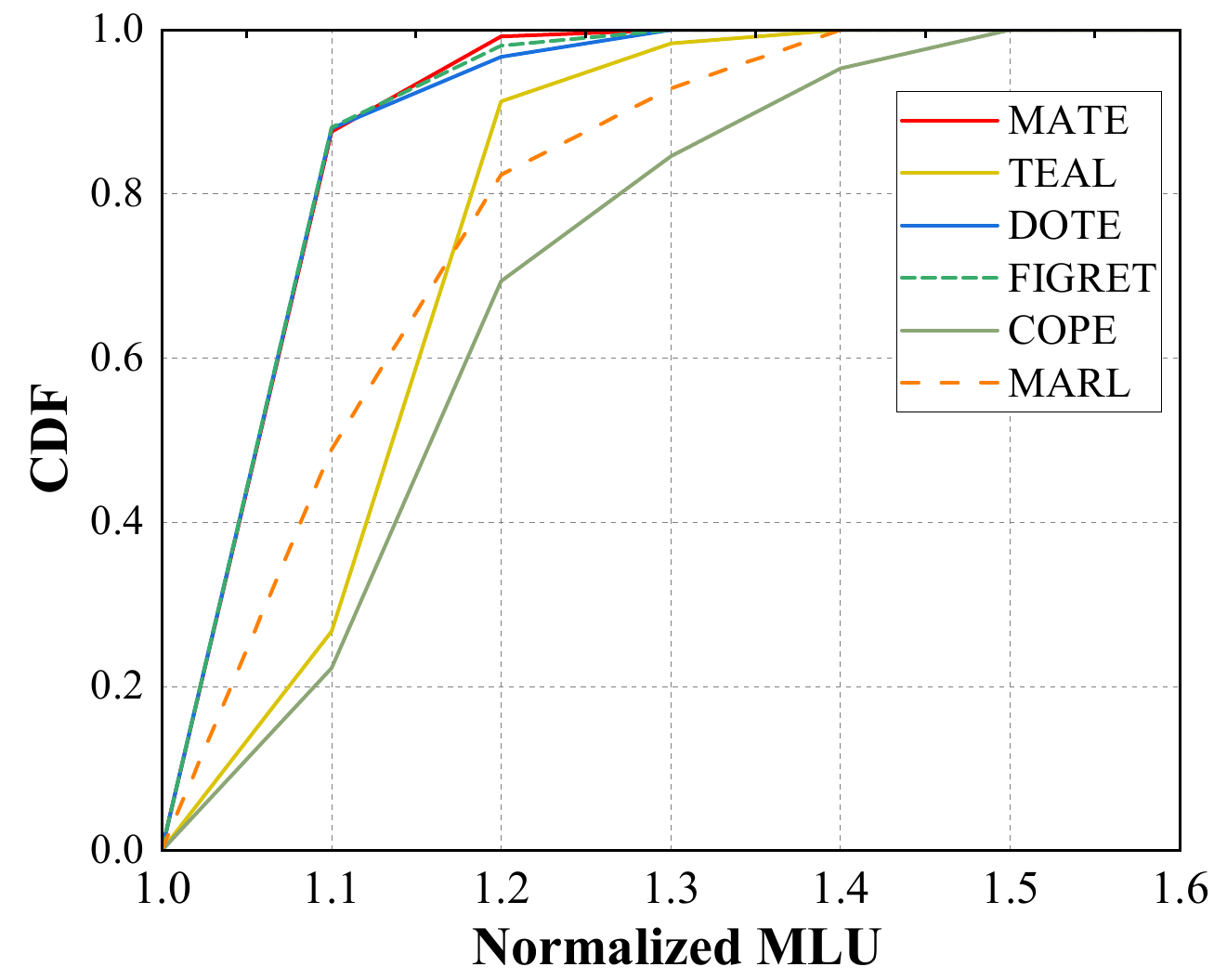}
        \caption{\small Deployment ratio: $1.0$}
    \end{subfigure}

    \caption{\small CDFs of normalized MLU on CENI-Bandcon under different SRv6 deployment ratios.}
    \label{bandcon前瞻性}
\end{figure*}

 Figs.~\ref{Abilene前瞻性}--\ref{bandcon前瞻性}  compare the cumulative distribution functions (CDFs) of normalized MLU on the CENI-Abilene, CENI-G\'EANT, and CENI-Bandcon testbed topologies across SRv6 deployment ratios.
Normalized MLU is measured relative to the sample-specific LP optimum under the corresponding evaluation configuration. 
The same reference is used for all compared methods, and values closer to one indicate performance closer to that optimum.
A higher CDF at a fixed threshold, or a leftward shift at a fixed quantile, indicates lower normalized MLU.

\par As shown in Fig.~\ref{Abilene前瞻性}, on CENI-Abilene, MATE achieves low normalized MLU at deployment ratios of $0.3$, $0.6$, and $1.0$, keeping it within $1.1$ for up to $90.1\%$ of samples.
The CDFs of MARL, COPE, and TEAL are generally shifted to the right, indicating higher link utilization across more samples. 
COPE provides worst-case protection against demand uncertainty, but its conservatism can impose unnecessary routing costs on stable demands.
TEAL uses the preceding interval's TM in our evaluation, so abrupt subsequent demand increases can overload links favored by its configuration.

\par As shown in Fig.~\ref{geant前瞻性}, on CENI-G\'EANT, the CDFs of MATE and FIGRET are generally close and outperform those of MARL, COPE, and TEAL.
At a deployment ratio of $0.3$, both MATE and FIGRET keep normalized MLU at approximately $1.2$ or below for about $90\%$ of the test samples.
The other three methods cover smaller fractions at the same threshold.
These results show that MATE achieves congestion-control performance close to FIGRET on measured traffic time series despite generating decisions locally rather than using network-wide traffic history to produce a global configuration.
Across the evaluated scenarios, MATE, DOTE, and FIGRET generally outperform TEAL and MARL. 
However, DOTE trails MATE and FIGRET on CENI-Abilene and CENI-G\'EANT, suggesting that history-based routing without explicit burst-specific protection can be insufficient under abrupt demand changes.
MATE and FIGRET use demand-variability-dependent penalties to protect burst-prone demands.
MATE additionally uses local topology features and distributed execution, maintaining comparable congestion control with less reliance on centralized online policy generation.

\par As shown in Fig.~\ref{bandcon前瞻性}, on CENI-Bandcon, the performance differences among MATE, DOTE, and FIGRET are smaller.
This smaller gap may reflect synthetic TMs that do not capture the full burstiness and irregularity of measured traffic, allowing DOTE's history-based policy to remain competitive.
Together with the results on CENI-Abilene and CENI-G\'EANT, this observation suggests greater benefits from explicit burst protection
on the measured traces.

\par Overall, MATE outperforms MARL, COPE, and TEAL in most evaluated scenarios and remains close to FIGRET without centralized online optimization.
Its design combines stability-aware protection for heterogeneous demands, capacity-aware path-sensitivity control, and counterfactual credit assignment that uses a shared global critic to guide agent-specific policy learning.

\vspace{-1em}

\subsection{Simulation Evaluation}\label{2仿真}

\par We further evaluate MATE on Sanet from TopologyZoo~\cite{topozoo}, whose 43 nodes and 90 links exceed the sizes of all three testbed topologies.
The OpenAI Gym-based environment~\cite{gym} runs on an Intel Core i9-12900H machine with Ubuntu~22.04 LTS, and the remaining settings follow the WAN testbed evaluation.

\begin{figure*}[tb]
    \centering
    \begin{subfigure}[b]{0.31\linewidth}
        \centering
        \includegraphics[width=\linewidth]
        {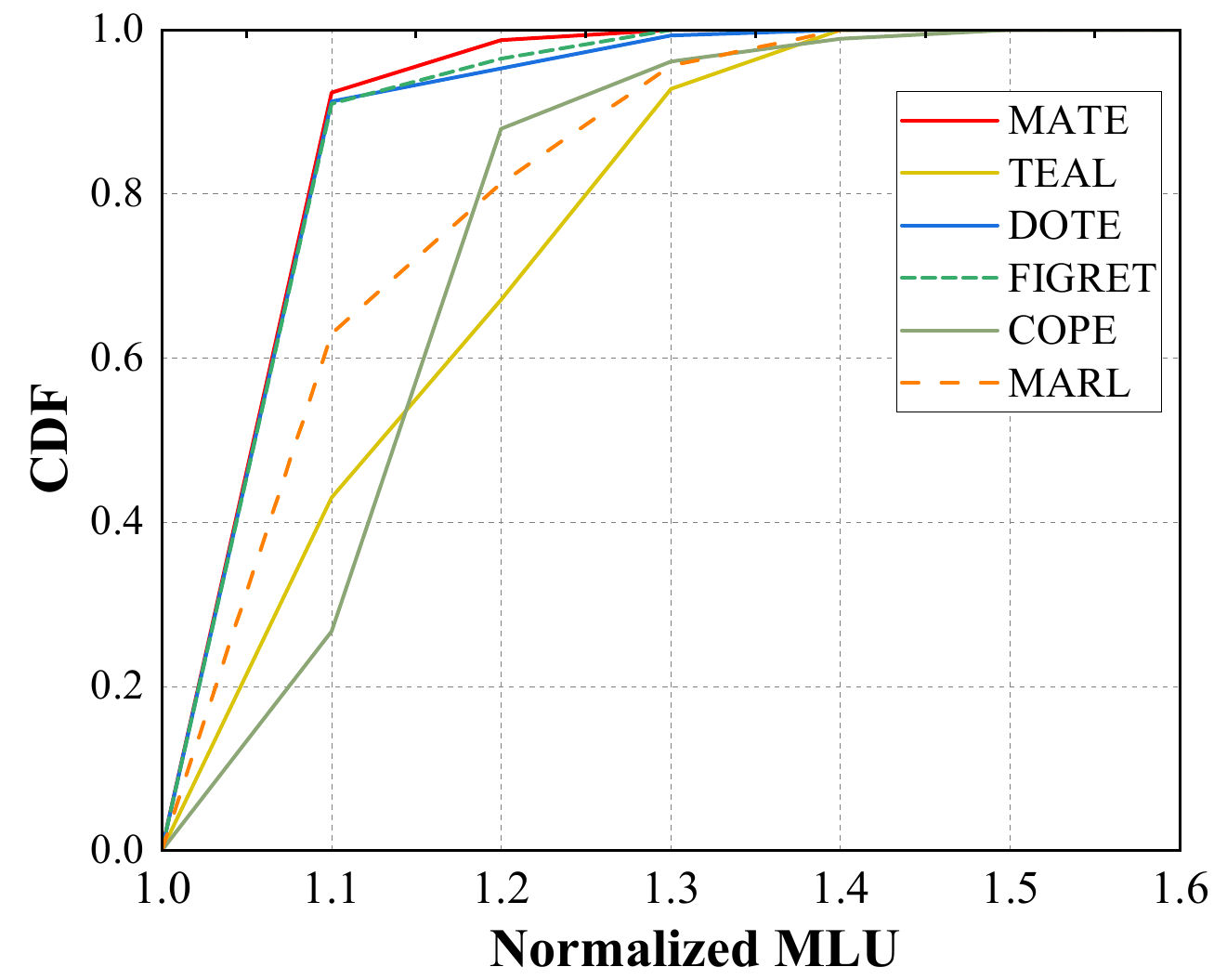}
        \caption{\small Deployment ratio: $0.3$}
    \end{subfigure}
    \hfill
    \begin{subfigure}[b]{0.31\linewidth}
        \centering
        \includegraphics[width=\linewidth]
        {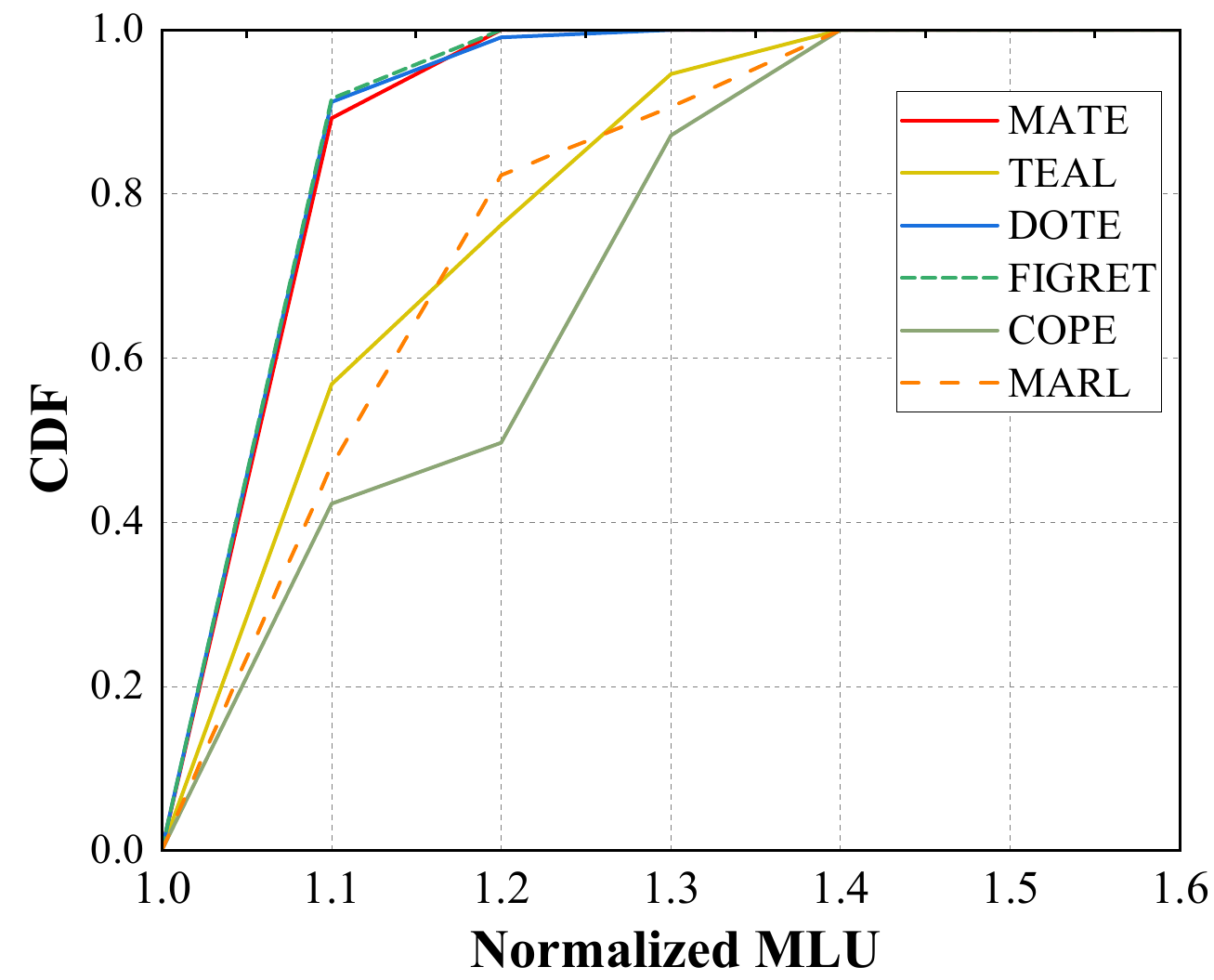}
        \caption{\small Deployment ratio: $0.6$}
    \end{subfigure}
    \hfill
    \begin{subfigure}[b]{0.31\linewidth}
        \centering
        \includegraphics[width=\linewidth]
        {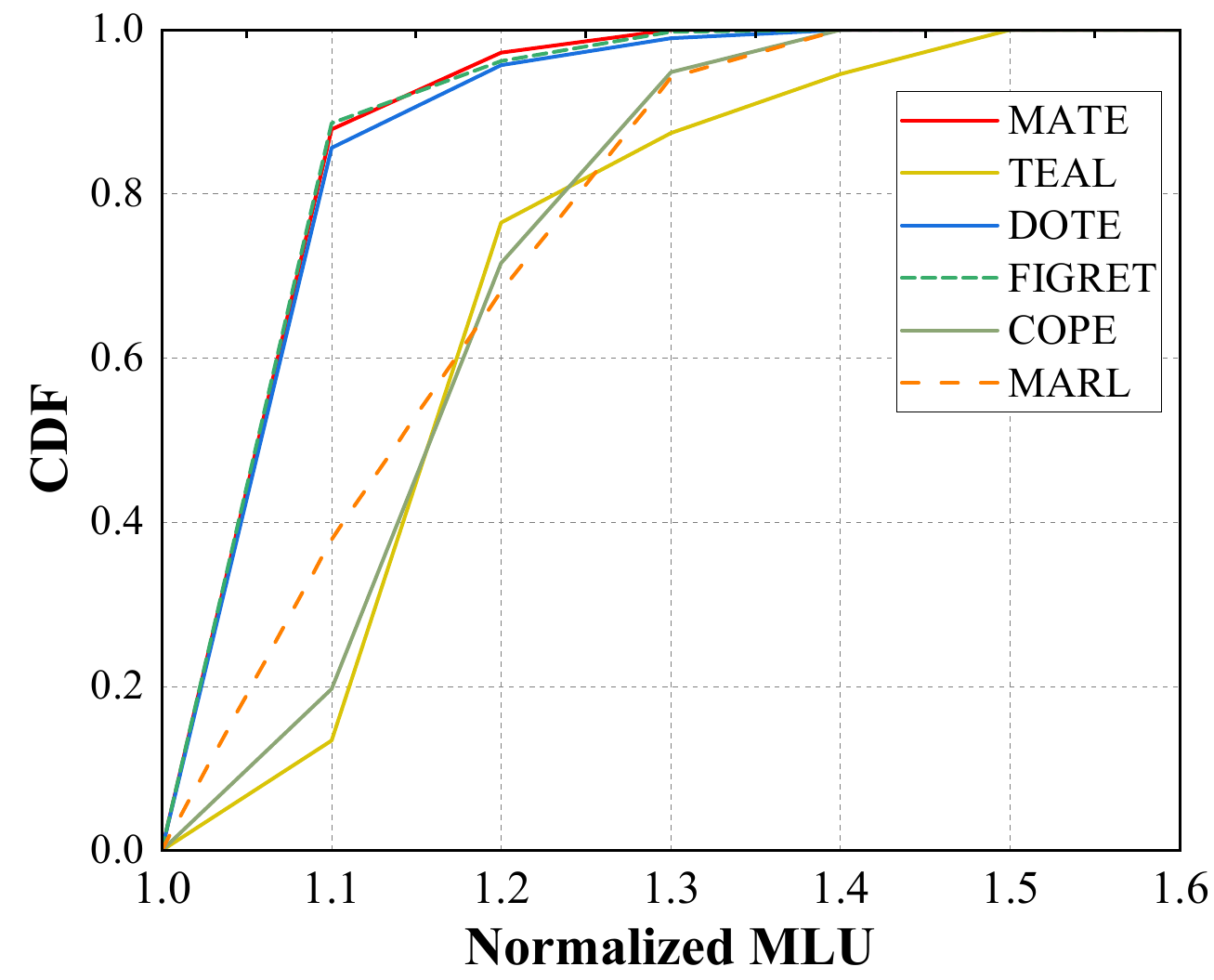}
        \caption{\small Deployment ratio: $1.0$}
    \end{subfigure}
    \caption{\small CDFs of normalized MLU on Sanet under different SRv6 deployment ratios.}
    \label{sanet前瞻性}
    \vspace{-1em}
\end{figure*}

\par As shown in Fig.~\ref{sanet前瞻性}, MATE maintains low normalized MLU across the evaluated SRv6 deployment ratios.
At a deployment ratio of $0.3$, MATE, FIGRET, and DOTE keep normalized MLU near or below $1.1$ for a large fraction of test samples, whereas the CDFs of MARL, COPE, and TEAL are generally shifted to the right. 
At deployment ratios of $0.6$ and $1.0$, MATE remains close to FIGRET and DOTE and generally outperforms the other three methods.
These results support MATE's applicability to the larger tested network.
The Sanet evaluation uses synthetic TMs because measured time series are unavailable for these experiments.
Therefore, the smaller performance gap among DOTE, MATE, and FIGRET may reflect the synthetic traffic characteristics, rather than equivalent performance under measured bursts.
The CENI-Abilene and CENI-G\'EANT experiments address measured traffic, whereas Sanet establishes applicability at the larger tested scale.

\vspace{-1em}

\subsection{Performance Under Link Failures}

\par Link failures reduce available network resources, invalidate candidate paths, and shift traffic onto surviving links.
Given the strong and comparable performance of MATE and FIGRET in Sec. \ref{2真实} and Sec. \ref{2仿真}, we compare the two methods under different numbers of link failures.
For each $k\in\{2,4,6\}$, we generate 10 scenarios by randomly removing $k$ bidirectional links while preserving connectivity, and evaluate both methods on identical test traffic.
For CENI-Abilene, we consider only $k\in\{2,4\}$, since further failures leave too few candidate paths for meaningful comparison.
The LP reference is recomputed for the surviving topology and the corresponding feasible candidate paths, and the same reference is used to normalize both methods.

\begin{figure*}[t]
    \centering
    \begin{subfigure}[b]{0.31\linewidth}
        \centering
        \includegraphics[width=\linewidth]
        {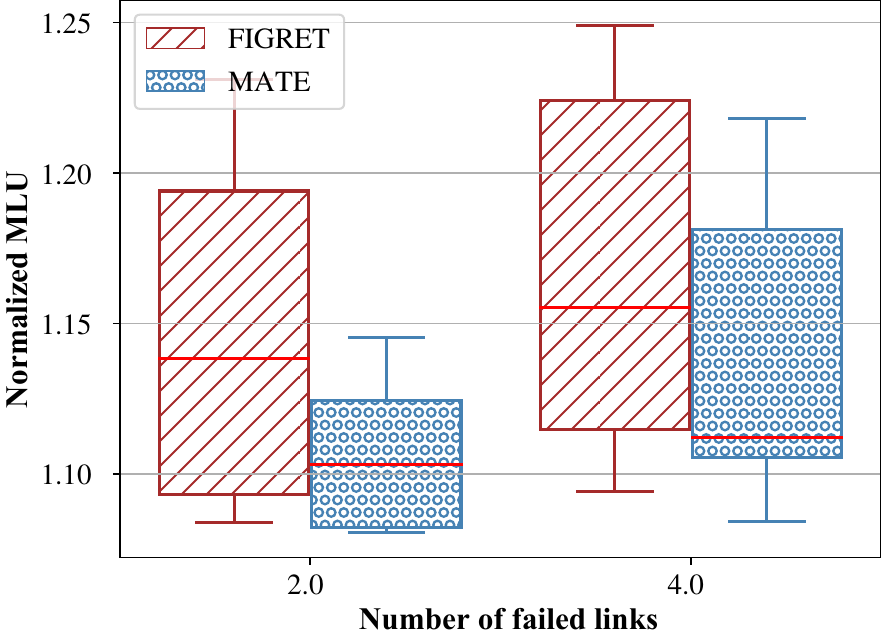}
        \caption{\small CENI-Abilene}
    \end{subfigure}
    \hfill
    \begin{subfigure}[b]{0.31\linewidth}
        \centering
        \includegraphics[width=\linewidth]
        {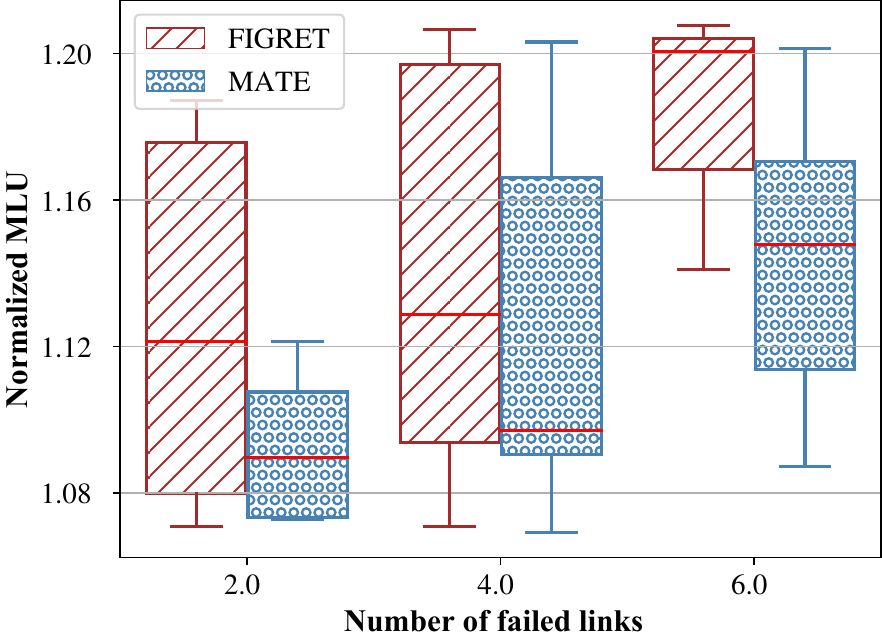}
        \caption{\small CENI-G\'EANT}
    \end{subfigure}
    \hfill
    \begin{subfigure}[b]{0.31\linewidth}
        \centering
        \includegraphics[width=\linewidth]
        {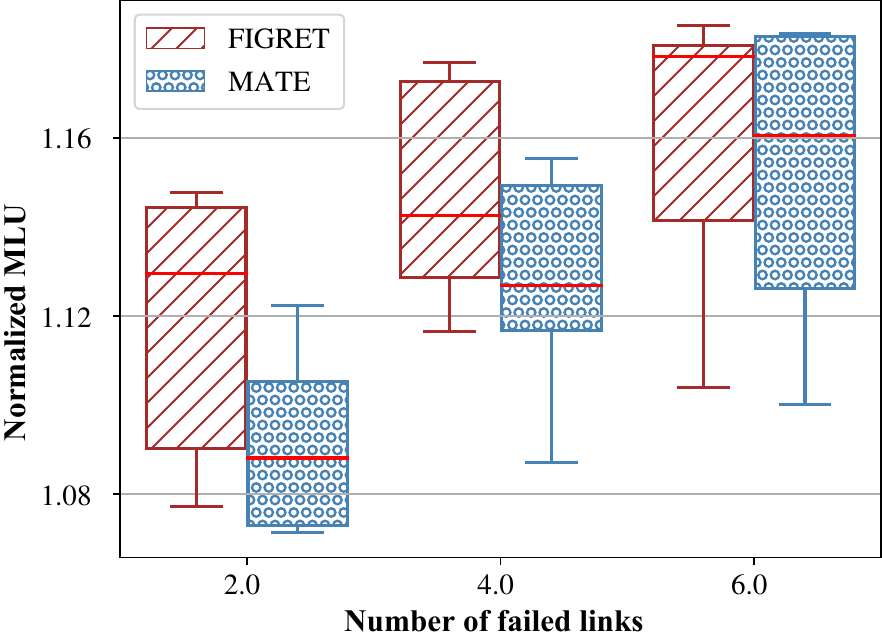}
        \caption{\small CENI-Bandcon}
    \end{subfigure}
    \caption{\small Performance of MATE and FIGRET under different numbers of link failures.}
    \label{2故障}
    \vspace{-1em}
\end{figure*}

\par As shown in Fig.~\ref{2故障}, normalized MLU generally increases with the number of failed links, but MATE has lower medians than FIGRET at most evaluated failure levels on CENI-Abilene, CENI-G\'EANT, and CENI-Bandcon.
On CENI-Abilene, FIGRET and MATE have respective medians of approximately $1.14$ and $1.10$ with two failed links, and $1.16$ and $1.11$--$1.12$ with four, yielding roughly $3\%$--$4\%$ reductions.
On CENI-G\'EANT, MATE reduces median normalized MLU by approximately $1\%$--$5\%$ relative to FIGRET across the evaluated failure levels.
MATE also maintains lower median values on CENI-Bandcon.
These results are consistent with benefits from topology-aware inputs: MATE conditions decisions on locally observed link utilization and capacity.
FIGRET redistributes traffic from failed paths through a separate recovery mechanism~\cite{figret}, but its neural policy does not explicitly take current topology features as inputs.
The comparison suggests that link-state conditioning can improve robustness as failures alter available paths and contention on surviving links.

\vspace{-1em}

\subsection{Effects of Candidate Paths and Neural Network Architecture}
\label{相关因素}

\par
\begin{figure}[h]
    \centering

    \begin{subfigure}[b]{0.48\linewidth}
        \centering
        \includegraphics[width=\linewidth]
        {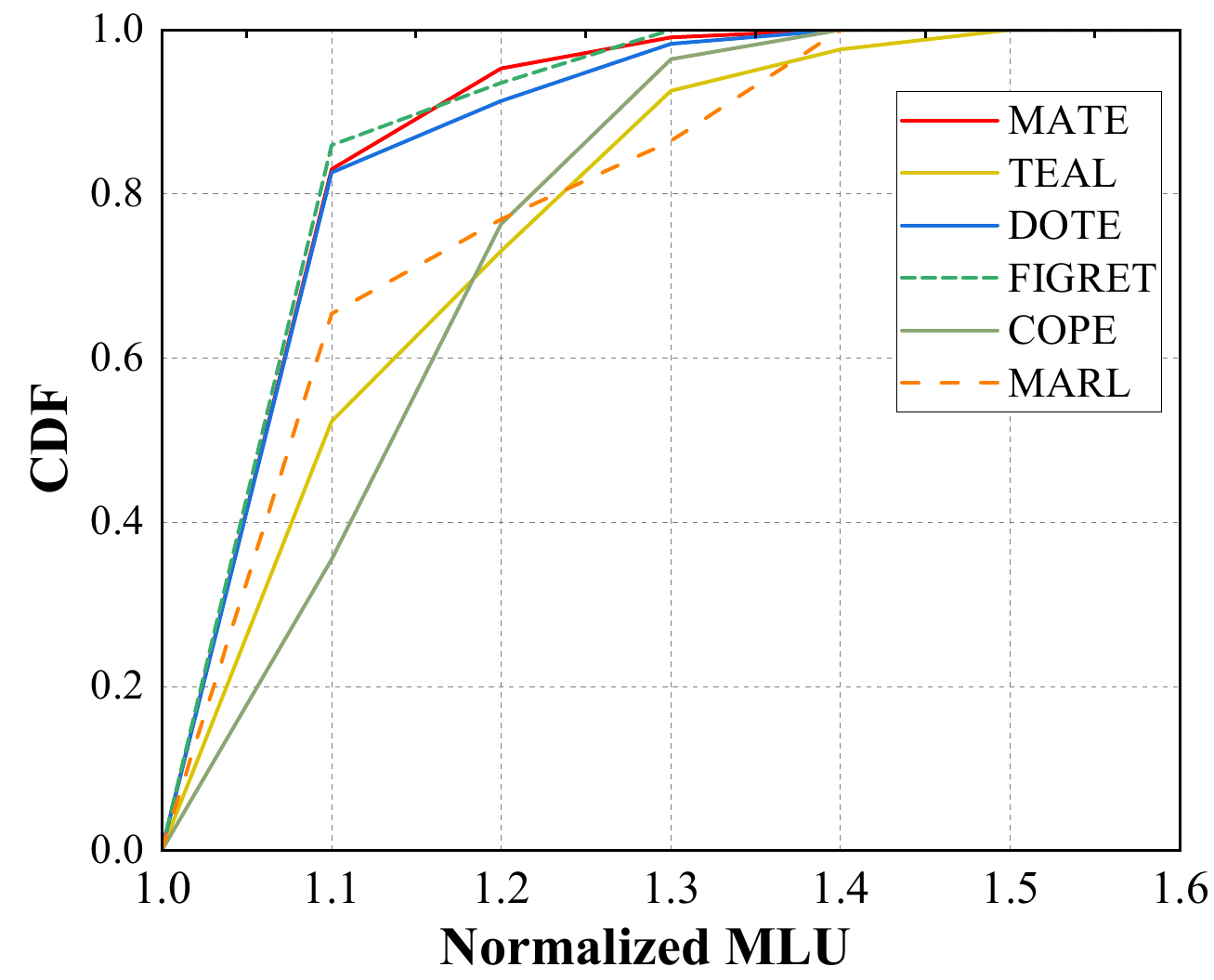}
        \caption{\small CENI-Abilene, ratio $0.3$}
    \end{subfigure}
    \hfill
    \begin{subfigure}[b]{0.48\linewidth}
        \centering
        \includegraphics[width=\linewidth]
        {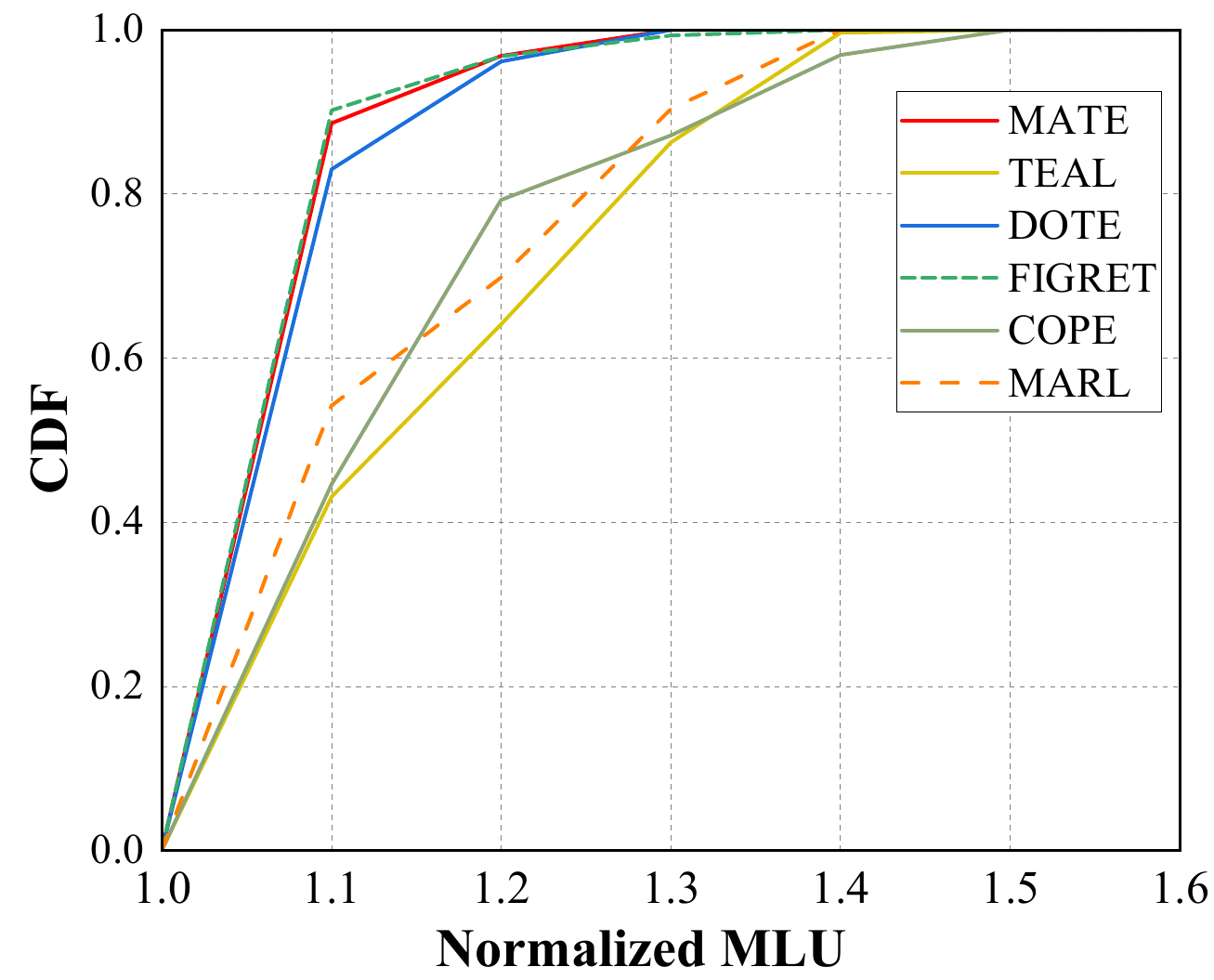}
        \caption{\small CENI-Abilene, ratio $1.0$}
    \end{subfigure}

    \par\smallskip

    \begin{subfigure}[b]{0.48\linewidth}
        \centering
        \includegraphics[width=\linewidth]
        {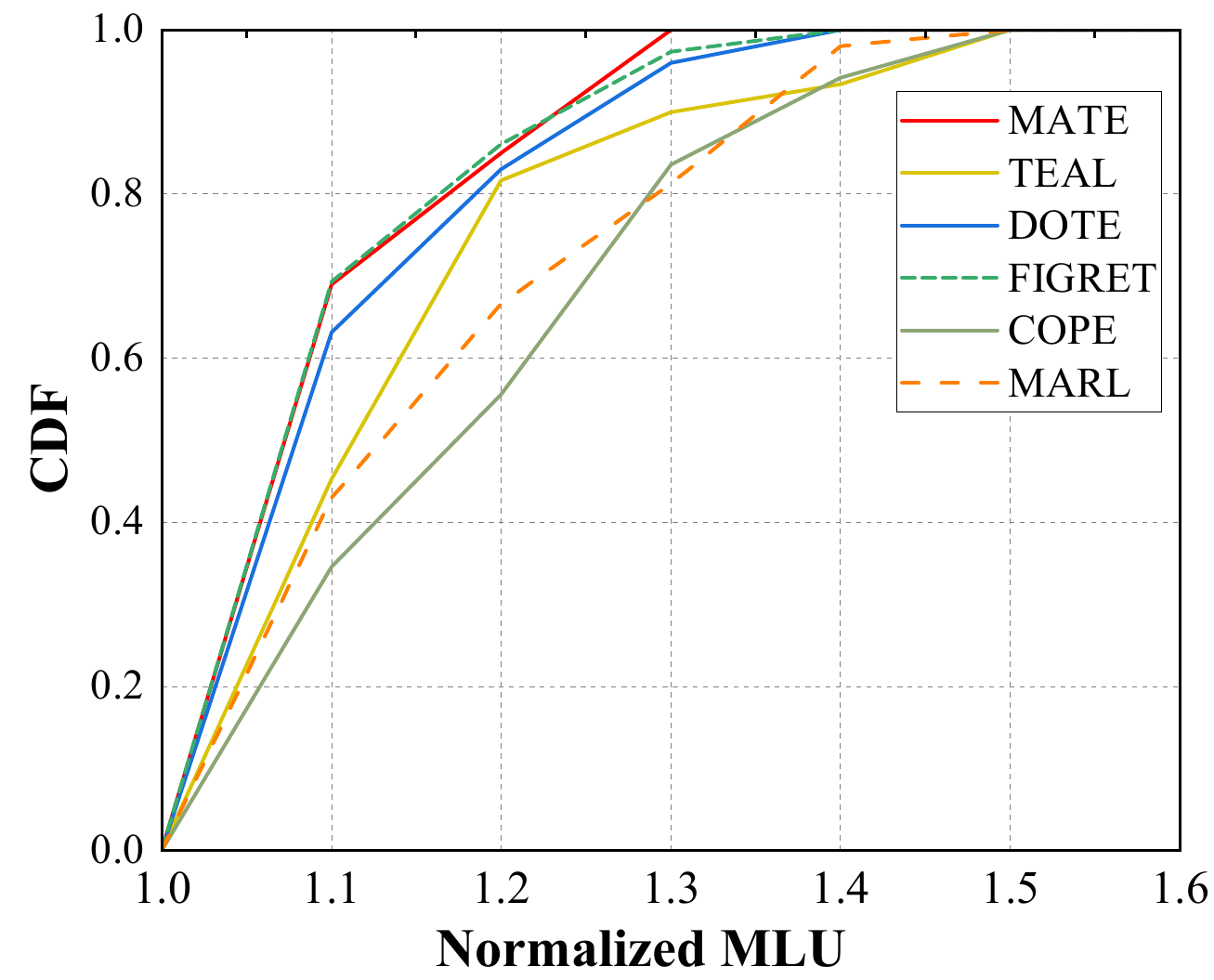}
        \caption{\small CENI-G\'EANT, ratio $0.3$}
    \end{subfigure}
    \hfill
    \begin{subfigure}[b]{0.48\linewidth}
        \centering
        \includegraphics[width=\linewidth]
        {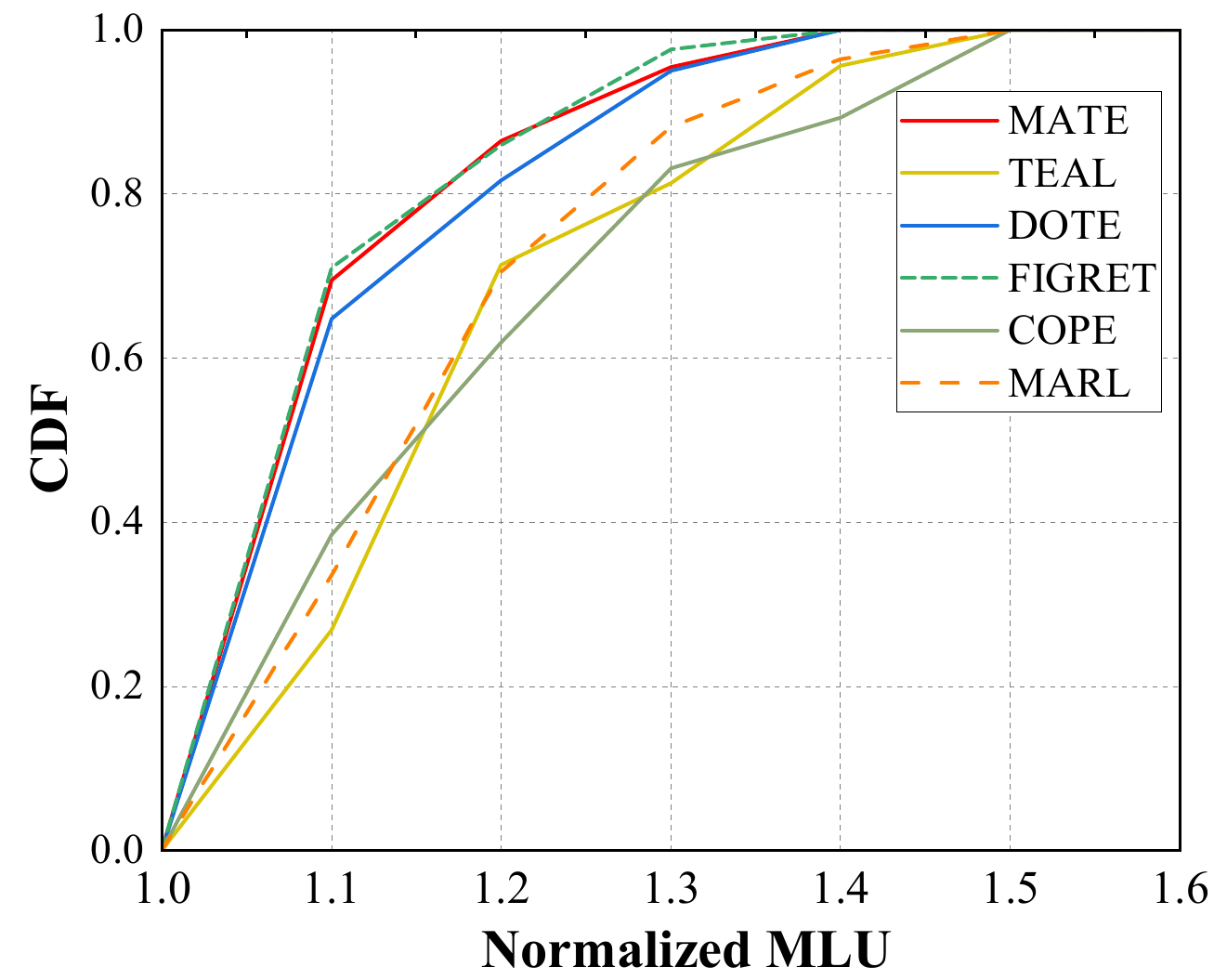}
        \caption{\small CENI-G\'EANT, ratio $1.0$}
    \end{subfigure}
    \caption{\small CDFs of normalized MLU with R\"acke-based candidate path construction.}
    \label{2路径}
    \vspace{-1.5em}
\end{figure}

 We assess MATE's sensitivity to candidate-path construction by replacing $K$-shortest paths with R\"acke-based paths, following semi-oblivious TE~\cite{semi}, under the same SRv6 deployment and path-realizability constraints. The LP reference is recomputed using the corresponding candidate-path set, with the remaining constraints unchanged.
The underlying oblivious routing framework bounds worst-case congestion relative to optimal demand-aware routing across arbitrary demands.
Fig.~\ref{2路径} presents the resulting normalized MLU CDFs on CENI-Abilene and CENI-G\'EANT under different SRv6 deployment ratios. 
Compared with Fig.~\ref{Abilene前瞻性} and Fig.~\ref{geant前瞻性}, MATE's performance changes only modestly under the alternative path construction.
MATE, FIGRET, and DOTE continue to outperform COPE, TEAL, and MARL, indicating robustness to the tested path constructions.
Changing the path construction alone does not eliminate the performance gap of the other three methods under the evaluated traffic conditions. Next, we examine sensitivity to neural network architecture on CENI-Abilene at a deployment ratio of $0.3$.
The default actor and critic architectures are listed in Table~\ref{para2}. Table~\ref{神经网络} examines alternative numbers and widths of their hidden layers.
Mean normalized MLU ranges from $1.052$ to $1.060$, a relative variation below $1\%$, indicating limited sensitivity to the tested architectures and allowing flexible neural network configuration.

\begin{table}[h]
    \centering
    \caption{\small Effect of Hidden-Layer Configurations on MATE Performance on CENI-Abilene}
    \label{神经网络}
    \footnotesize
    \renewcommand{\arraystretch}{1.15}
    \setlength{\tabcolsep}{4pt}

    \begin{tabular}{@{}ccc@{}}
        \toprule
        Actor widths & Critic widths & Mean normalized MLU \\
        \midrule
        $(128,64,32)$ & $(128,64,64)$ & $1.054$ \\
        $(128,32,32)$ & $(128,32,64)$ & $1.052$ \\
        $(128,64)$    & $(128,64)$    & $1.056$ \\
        $(128,64)$    & $(64,64)$     & $1.060$ \\
        \bottomrule
    \end{tabular}
\end{table}

\vspace{-1.5em}

\subsection{Online Inference Time}\label{2计算时间}

\par Table~\ref{2时间} compares mean online inference times. 
MATE requires $2.05$--$3.14$~ms and is faster than FIGRET on all three topologies, with reductions of $22.3\%$, $9.8\%$, and $6.3\%$ on CENI-Abilene, CENI-G\'EANT, and CENI-Bandcon, respectively, while remaining comparable to MARL.
Despite incorporating topology features, MATE retains low latency through node-level policy inference, supporting the computational efficiency of distributed scheduling.
These measurements exclude the remaining online control-loop stages.

\begin{table}[h]
    \centering
    \caption{\small Mean Online Inference Time (ms)}
    \label{2时间}
    \renewcommand{\arraystretch}{1.1}

    \begin{tabular}{@{}lccc@{}}
        \toprule
        Topology & MATE & FIGRET & MARL \\
        \midrule
        CENI-Abilene & $2.05$ & $2.64$ & $2.32$ \\
        CENI-G\'EANT & $2.85$ & $3.16$ & $2.96$ \\
        CENI-Bandcon & $3.14$ & $3.35$ & $3.08$ \\
        \bottomrule
    \end{tabular}
\end{table}

\vspace{-1em}

\section{Conclusion and Future Outlook}

\par In this paper, we proposed MATE for proactive TE in hybrid IP/SRv6 networks. 
MATE combines traffic stability with capacity-aware path sensitivity to protect burst-prone demands while limiting conservatism for stable traffic.
MATE augments a MADDPG-based centralized training framework with shared-critic counterfactual credit assignment, using agent-specific difference rewards to guide local policy learning toward the network-wide congestion-and-sensitivity objective.
Online execution relies only on local observations.
Across SRv6 deployment ratios, CENI-WAN experiments show congestion control close to centralized FIGRET on measured traffic.
On CENI-Abilene, MATE converges after approximately 500 episodes, reduces converged normalized MLU by approximately $21\%$ relative to MARL, and keeps normalized MLU within $1.1$
for up to $90.1\%$ of test instances.
Under tested link failures on CENI-G\'EANT, it reduces median normalized MLU by $1\%$--$5\%$ relative to FIGRET.
Online inference takes $2.05$--$3.14$~ms, up to $22.3\%$ less time than FIGRET.
These results support proactive TE scheduling from local observations with reduced dependence on centralized online control.

\par Future work will investigate lightweight adaptation to unseen topologies and evolving traffic distributions, and will evaluate MATE in larger operational networks with more diverse failure patterns. 
Beyond neural inference time, we will measure the complete control loop, including observation collection, policy application, and failure recovery, to quantify how deployment overhead affects the trade-off between distributed responsiveness and global congestion control.

\footnotesize
\bibliographystyle{IEEEtranN}
\bibliography{IEEEabrv,ref}

\vspace{-0.8cm}

\end{document}